\documentclass[sigconf]{acmart}
\AtBeginDocument{%
  }
    
\usepackage{multirow}
\usepackage{makecell}
\usepackage[normalem]{ulem}
\useunder{\uline}{\ul}{}
\usepackage{placeins}
\usepackage{enumitem}
\usepackage{algorithmic}
\usepackage[ruled]{algorithm2e}

\setcopyright{acmlicensed}
\copyrightyear{2018}
\acmYear{2018}
\acmDOI{XXXXXXX.XXXXXXX}
\acmConference[Conference acronym 'XX]{Make sure to enter the correct
  conference title from your rights confirmation email}{June 03--05,
  2018}{Woodstock, NY}
\acmISBN{978-1-4503-XXXX-X/2018/06}

\begin{document}

\title{TimeRFT: Stimulating Generalizable Time Series Forecasting for TSFMs via Reinforcement Finetuning}

\author{Siyang Li}
\affiliation{%
  \institution{HKUST(GZ)}
  \city{Guangzhou}
  \country{China}}
\email{sli572@connect.hkust-gz.edu.cn}

\author{Yize Chen}
\affiliation{%
  \institution{University of Alberta}
  \state{Edmonton}
  \country{Canada}
}
\email{yize.chen@ualberta.ca}

\author{Zijie Zhu}
\affiliation{%
  \institution{Alibaba Cloud}
  \city{Guangzhou}
  \country{China}
}
\email{zhuzijie.zzj@alibaba-inc.com}

\author{Yuxin Pan}
\affiliation{%
  \institution{City University of Hong Kong}
  \city{Hong Kong}
  \country{China}
}
\email{yuxin.pan@connect.ust.hk}

\author{Yan Guo}
\affiliation{%
  \institution{Alibaba Cloud}
  \city{Hangzhou}
  \country{China}
}
\email{qingjian.gy@alibaba-inc.com}

\author{Ming Huang}
\affiliation{%
  \institution{Alibaba Cloud}
  \city{Guangzhou}
  \country{China}
}
\email{mingqian.hm@alibaba-inc.com}

\author{Hui Xiong}
\authornote{Corresponding author.}
% \correspondingauthor
\affiliation{%
  \institution{HKUST(GZ) \& HKUST}
  \city{Guangzhou}
  \country{China}}
\email{xionghui@ust.hk}

\renewcommand{\shortauthors}{Trovato et al.}

\begin{abstract}
Time Series Foundation Models (TSFMs) have demonstrated strong generalization capability and data efficiency in time series forecasting through large-scale pretraining. However, adapting TSFMs to downstream forecasting tasks remains challenging due to temporal distribution shifts and varying data availability. Specifically, the non-stationary and uncertain nature of time series data leads to discrepancies between historical training and future forecasting distributions, making existing Supervised FineTuning (SFT)-based adaptation vulnerable to overfitting and limited generalization. Moreover, forecasting tasks often operate under varying data regimes, requiring TSFMs to extract generalizable temporal patterns from limited training samples. To address these challenges, we propose \underline{T}ime series \underline{R}einforcement \underline{F}ine\underline{T}uning (TimeRFT), a reinforcement learning-based adaptation paradigm for TSFMs. TimeRFT introduces two forecasting-oriented training recipes: (i) A quality-aware temporal reward mechanism providing fine-grained credit assignment by holistically evaluating the contribution of each prediction step to overall forecasting performance. (ii) A difficulty-aware data selection strategy prioritizing informative time series samples with generalizable forecasting patterns. Extensive experiments on diverse real-world forecasting benchmarks demonstrate that TimeRFT consistently surpasses SFT-based adaptation methods across various real-world forecasting tasks with different data regimes, achieving improved prediction accuracy and enhanced generalization against unforeseen distribution shifts. Our code is available at 
\url{https://github.com/LSY-Cython/TimeRFT}.
% \url{https://anonymous.4open.science/r/TimeRFT-5634}.
\end{abstract}



\maketitle

\section{Introduction}
Time series forecasting plays a crucial role in a wide range of real-world applications \cite{faloutsos2018forecasting}. Conventional deep learning-based methods \cite{wang2024deep, cui2021metro, li2025ufgtime} typically require training scenario-specific models, which limits their scalability and cross-scenario generalization capability. Time Series Foundation Models (TSFMs) have emerged as a promising paradigm by leveraging large-scale and heterogeneous time series pretraining \cite{liang2024foundation}. By learning universal temporal representations from diverse datasets, TSFMs achieve remarkable zero-shot forecasting capability across unseen scenarios \cite{aksu2024gift}. A growing number of domain-specific TSFMs have been developed to support data-driven decision-making in various fields, including energy \cite{tu2024powerpm}, healthcare \cite{li2025mira}, finance \cite{zhu2025fincast} and cloud computing \cite{xie2025chatts}.

Despite their strong zero-shot performance, effectively adapting TSFMs to downstream forecasting tasks remains challenging. Existing TSFM research primarily focuses on unified pretraining frameworks, architecture design, and large-scale data curation \cite{liu2024timer, ansari2024chronos, shi2025timemoe, woo2024unified, das2024decoder}, while task-specific finetuning is relatively underexplored. Several TSFMs \cite{liu2024timer, ekambaram2024tiny, shi2025timemoe, goswami2024moment} have reported few-shot or full-shot finetuning results on downstream datasets, but they still suffers from temporal distribution shifts between historical training and future forecasting scenarios. Recent parameter-efficient adaptation approaches \cite{chen2025visionts, zhao2025less, qiao2025multiscale, na2026timepeft} alleviate this issue by selectively updating task-related parameters, but their generalizability to unseen temporal patterns remains lacking. Moreover, existing TSFM adaptation methods are rarely evaluated under diverse training data regimes, such as few-shot and full-shot settings \cite{li2025tsfm, shchur2025fev}.

Most existing TSFM adaptation methods follow the Supervised FineTuning (SFT) paradigm, which optimizes TSFMs by directly imitating observed forecasting targets. However, SFT can easily overfit limited training sequences by exploiting spurious temporal correlations and random noise \cite{qiao2025multiscale, zhao2025less}. Besides, SFT typically treats all training samples equally, without identifying and removing low-quality time series containing non-generalizable predictive patterns \cite{fu2025selective, wu2025enhancing}. These two drawbacks render SFT-adapted TSFMs often struggle to generalize under temporal distribution shifts and unseen forecasting scenarios \cite{liu2024time}.

Reinforcement FineTuning (RFT) provides a promising alternative by enabling models to improve through self-exploration, rather than relying solely on ground-truth imitation like SFT. RFT learns from diverse self-generated samples, rich trial-and-error experience and task-specific rewards, which can potentially alleviate overfitting and improve generalization \cite{wu2025generalization, trung2024reft}. For Large Language Models (LLMs), recent advances have demonstrated effectiveness of RFT in enhancing model reasoning capabilities \cite{guo2025deepseek, team2025kimi, lambert2024tulu}. While directly applying RFT to TSFM adaptation remains challenging due to unique characteristics of time series forecasting.

Specifically, two fundamental challenges need to be addressed. The first challenge is \textit{how to achieve reliable step-wise credit assignment for forecasting trajectories}. Fine-grained credit assignment is crucial for identifying which prediction steps contribute positively to the final forecasting quality and for guiding stable policy optimization. In LLM reasoning, existing approaches \cite{lightman2023let, setlur2025rewarding, zhou2025sequence} typically introduce auxiliary neural reward models to evaluate intermediate reasoning steps. However, such process-based reward modeling suffers from reward hacking and requires expensive human annotations, often achieving inferior performance compared with outcome-based rewards \cite{gao2024designing, guo2025deepseek, havrilla2024teaching}. In contrast, time series forecasting naturally provides step-wise ground-truth observations. Therefore, the key challenge is \textit{how to effectively exploit available temporal supervision to design dense, reliable, and forecasting-oriented rewards that accurately evaluate the contribution of each prediction step}.

The second challenge is \textit{how to select informative time series samples for effective RFT over TSFMs}. As real-world time series data often contain noise, anomalies and irregular temporal patterns, training data quality plays a critical role in both forecasting and reinforcement learning (RL). Caveats in data samples may introduce non-generalizable correlations and diminish sequence forecastability, leading to degraded generalization performance \cite{yang2025not, fu2025selective, wu2025enhancing}. Meanwhile, we observe that RFT benefits from training samples of appropriate difficulty matching the model capability, since overly easy samples narrow the exploration space and tend to be excessively exploited towards achieving high rewards, while overly difficult samples lead to meaningless rewards and unstable policy optimization \cite{shi2025efficient, sun2025improving, li2025limr}. Therefore, \textit{accurately evaluating forecasting difficulty and selecting informative training sequences} remain essential for effective RFT-based TSFM adaptation.

To this end, we propose \textbf{TimeRFT}, a \underline{T}ime \underline{S}eries \underline{R}einforcement \underline{F}ine\underline{T}uning method that enables accurate and generalizable adaptation of TSFMs. TimeRFT specifically develops two forecasting-oriented training recipes for time series RL: \textbf{(i) Forecasting quality-based hybrid reward design.} It performs step-wise and multi-faceted evaluation of on-policy forecasting trajectories by jointly considering prediction accuracy and temporal structure alignment. Such dense reward signals provide reliable fine-grained credit assignment and encourage TSFMs to learn more generalizable temporal modes. \textbf{(ii) Forecasting difficulty-based data selection strategy.} It quantifies the forecasting difficulty of training samples based on the zero-shot performance of pretrained TSFMs and selects informative sequences with suitable forecastability, thereby improving exploration efficiency and stabilizing RFT training. Extensive experiments across diverse forecasting tasks and training data availability validate that TimeRFT achieves superior forecasting accuracy and generalization over SFT-based adaptation methods. The major contributions of this work are summarized as follows:
\begin{itemize}[leftmargin=*]
\item We propose a self-exploratory reinforcement learning paradigm TimeRFT for TSFM finetuning, which grounds forecasting quality into reward design and improves both accuracy and generalization under temporal distribution shifts.
\vspace{-1mm}
\item We introduce two forecasting-oriented RFT training strategies, including a quality-aware temporal reward mechanism for fine-grained credit assignment and a difficulty-aware training data selection for effective exploration.

\item We conduct comprehensive experiments across diverse real-world forecasting tasks and data regimes, demonstrating the state-of-the-art generalizable forecasting performance of TimeRFT over existing SFT-based approaches.
\end{itemize}

\section{Related Work}
\textbf{Time Series Foundation Models.}
Recent advances in large-scale pretraining have established TSFMs as a promising paradigm for general-purpose time series forecasting. Unlike conventional deep forecasting models \cite{liu2024itransformer, nie2023a, wang2024timemixer, qiu2025duet, chen2024similarity} that are typically designed for specific datasets or forecasting scenarios, TSFMs leverage heterogeneous time series pretraining to learn transferable temporal representations, enabling zero-shot forecasting, few-shot adaptation and cross-scenario generalization across diverse downstream tasks. Existing TSFM studies mainly focus on improving pretraining architectures, objectives and data scaling. Representative approaches explore unified architectures for handling heterogeneous temporal patterns \cite{liu2024timer, ansari2024chronos, auer2025tirex}, self-supervised forecasting objectives \cite{shi2025timemoe, liu2025moirai, liu2025sundial}, and large-scale time series corpus curation \cite{aksu2024gift, shao2025blast, liu2025empowering}. For example, Timer \cite{liu2024timer} adopts a decoder-only Transformer pretrained on large-scale time series data with a generative objective. Time-MoE \cite{shi2025timemoe} introduces a scalable mixture-of-experts architecture with multi-resolution forecasting objectives. MOIRAI \cite{woo2024unified} further explores unified probabilistic forecasting through large-scale pretraining on diverse temporal distributions.

Despite substantial progress in TSFM pretraining, downstream adaptation remains relatively underexplored. Existing studies mainly report naive finetuning results on individual datasets \cite{liu2024timer, shi2025timemoe, ekambaram2024tiny, rasul2023lagllama, auer2025tirex}, while recent parameter-efficient methods attempt to adapt task-specific temporal representations by updating a subset of model parameters \cite{qiao2025multiscale, zhao2025less, chen2025visionts, na2026timepeft}. However, most existing approaches still follow the SFT paradigm, which optimizes over observed training distributions and may suffer from overfitting under temporal distribution shifts and limited data regimes. In this work, we explore RFT as a new adaptation paradigm to improve the generalization capability of TSFMs for diverse forecasting scenarios.

\vspace{2pt}
\noindent \textbf{Reinforcement Learning with Verifiable Rewards (RLVR).} The success of recent LLMs such as OpenAI o1 \cite{jaech2024openai} and DeepSeek-R1 \cite{guo2025deepseek} has demonstrated the effectiveness of RL in enhancing complex reasoning capabilities. RLVR \cite{lambert2024tulu, guo2025deepseek} has been widely adopted to further improve LLM performance on tasks with automatically verifiable outcomes, including mathematical reasoning \cite{shao2024deepseekmath, yang2024qwen2}, code generation \cite{guo2024deepseek, hui2024qwen2} and robotic manipulation \cite{li2025simplevla, liu2025what}. To improve RLVR training stability and sample efficiency, various policy optimization algorithms have been proposed, including GRPO-based methods \cite{shao2024deepseekmath, zheng2025group, liu2025understanding, liu2026gdpo} and PPO variants \cite{kazemnejad2025vineppo}.

Although RLVR has demonstrated strong generalization and data efficiency in LLM reasoning, its potential for TSFM adaptation remains largely unexplored. Time series forecasting naturally provides verifiable prediction targets, making it a suitable domain for reward-driven optimization. Existing attempts \cite{luo2025time, niu2025langtime} mainly apply RL-based approaches to time series tasks through general-purpose LLMs rather than specialized TSFMs. \cite{qi2025timehf} pioneers PPO-based RL for aligning TSFM with human feedback, but it heavily relies on human preference annotation and does not full explore the generalizability of time series RL against diverse distribution shifts. In contrast, this work investigates the unique challenges of applying RLVR to TSFM finetuning, and develops forecasting-oriented reward design and data selection strategies to unlock the generalization capability of RFT for time series forecasting.

\begin{figure*}[ht]
\centering
\includegraphics[width=0.96\linewidth]{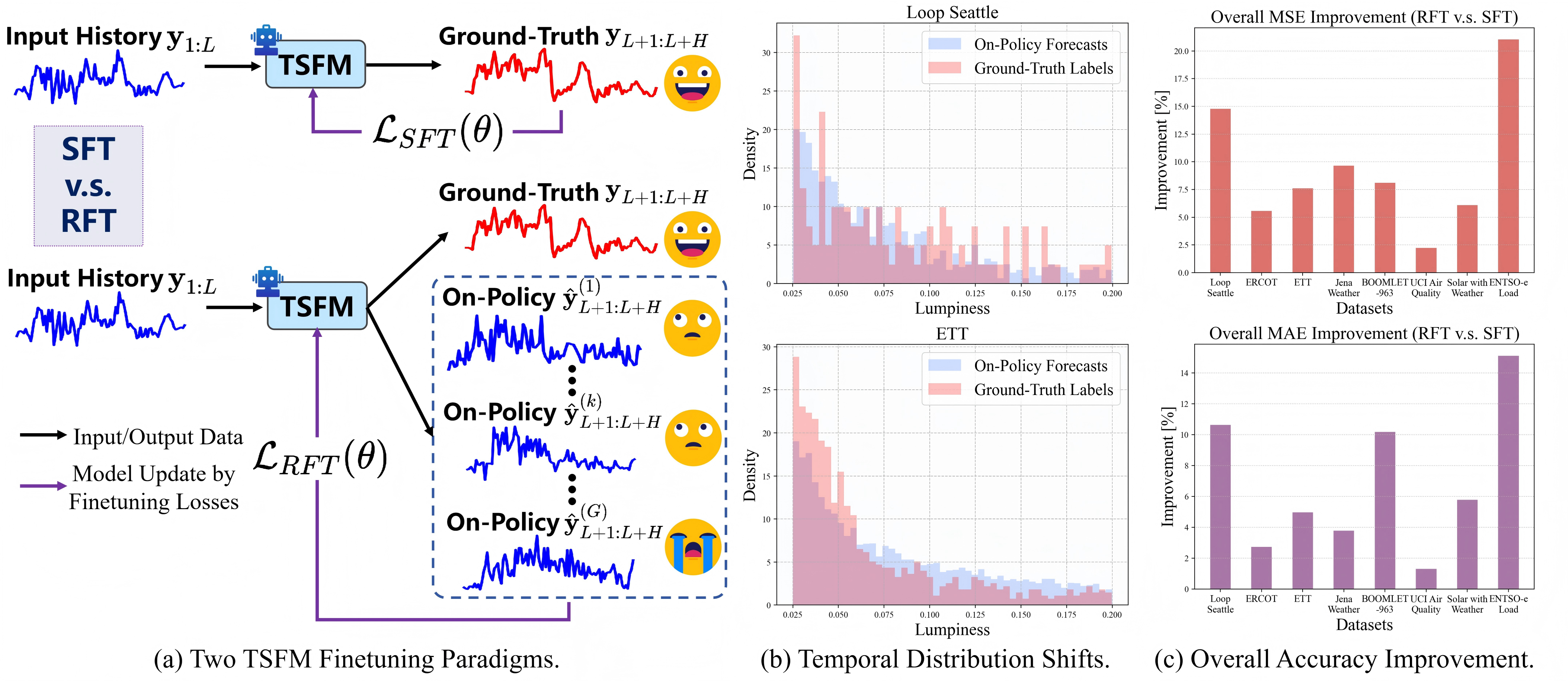}
\vspace{-6pt}
\caption{(a) Comparison between SFT and RFT for TSFM adaptation. (b) The lumpiness metric \cite{aksu2024gift} is adopted to characterize the distribution of temporal variability patterns across training time series samples used for on-policy RFT and imitative SFT. (c) Forecasting performance improvement of RFT over SFT.}
\label{fig:ft_paradigms}
\end{figure*}
\section{TSFM Finetuning Paradigms}
In this section, we formulate the downstream adaptation problem of TSFMs and introduce two representative finetuning paradigms. Figure \ref{fig:ft_paradigms} provides an overview of their learning mechanisms and illustrates the key differences in terms of training distribution and forecasting generalization.

\textbf{Problem Formulation.} 
Given a downstream time series dataset $\mathcal{D}=\{(\mathbf{x}_{i}, \mathbf{y}_{i})\}_{i=1}^{T}$, where $\mathbf{y}_{i} \in \mathbb{R}^{N_{d}}$ denotes the target series with $N_{d}$ variates, $\mathbf{x}_{i} \in \mathbb{R}^{N_{c}}$ denotes the observed covariates with $N_{c}$ channels, and $T$ denotes the total length of time series, the goal of TSFM adaptation is to finetune a pretrained TSFM for accurate future forecasting by learning a conditional predictive distribution
$f_{\theta}(\hat{\mathbf{y}}_{L+1:L+H}| \mathbf{y}_{1:L},\mathbf{x}_{1:L})$, where $L$ and $H$ indicate lookback window and prediction horizon respectively, $\hat{\mathbf{y}}_{i} \in \mathbb{R}^{N_{d}}$ indicates a model forecast at each time step $i$, and $f_{\theta}(\cdot)$ represents a trainable TSFM.

However, effective TSFM adaptation remains challenging due to two inherent characteristics of real-world forecasting scenarios. First, \textit{unforeseen temporal distribution shifts} widely exist between historical training sequences and future forecasting data due to the non-stationarity and stochasticity of time series \cite{liu2024time, kim2022reversible}. Therefore, adapted TSFMs should capture transferable temporal patterns which can cope with the potential shifts in future scenarios. Second, \textit{varying data availability} across downstream tasks requires TSFM finetuning methods to remain effective under diverse data regimes, including zero-shot, few-shot, and full-shot settings \cite{liu2024timer, ekambaram2024tiny}. Consequently, the primary objective of TSFM finetuning is to achieve accurate and generalizable forecasting under diverse temporal distributions and limited training resources.

\noindent \textbf{Supervised TSFM Finetuning.} 
SFT is the dominant approach for adapting pretrained TSFMs to downstream forecasting tasks \cite{na2026timepeft, gupta2024beyond, chen2025visionts}. Given historical observations $\mathbf{y}_{1:L}$ and covariates $\mathbf{x}_{1:L}$, SFT maximizes the conditional log-likelihood of generating the ground-truth future sequence $\mathbf{y}_{L+1:L+H}$ by the following loss function:
\begin{equation}
\vspace{-1pt}
\centering
\mathcal{L}_{SFT}(\theta)=\mathbb{E}_{(\mathbf{x},\mathbf{y})\sim \mathcal{D}}[\frac{1}{H}\sum_{i=L+1}^{L+H}-\log f_{\theta}(\mathbf{y}_{i}|\mathbf{y}_{1:i-1},\mathbf{x}_{1:i-1})].
\label{eq:1}
\vspace{-1pt}
\end{equation}

Although SFT effectively adapts TSFMs to specific datasets, it only optimizes predictive likelihood over the observed training distribution. Such behavior may encourage TSFMs to memorize and overfit dataset-specific temporal patterns, leading to degraded generalizability when encountering unseen forecasting distributions. Previous studies \cite{ekambaram2024tiny, chen2025visionts, qiao2025multiscale} have shown that SFT-based methods can suffer from unstable performance under different data scales and temporal distribution shifts, motivating the exploration of alternative finetuning paradigms with stronger generalization capability.

\vspace{2pt}
\noindent \textbf{Reinforcement TSFM Finetuning.}
RFT enables models to self-evolve via reward-driven exploration rather than rigidly imitating training samples. Recent advances in RFT have exhibited its effectiveness in enhancing the reasoning capability and generalization of LLMs \cite{guo2025deepseek, lambert2024tulu, jaech2024openai} under novel and complex contexts. Inspired by these successes, RFT provides a promising direction for improving TSFM adaptation under temporal distribution shifts. The learning objective of RFT is to maximize the reward of generated forecasting trajectories while constraining the deviation from the reference model through KL regularization \cite{lambert2024tulu}:
\begin{equation}
\centering
\resizebox{0.44\textwidth}{!}{$\max_{f_{\theta}} \mathbb{E}_{\hat{\mathbf{y}}_{o}\sim f_{\theta}(\hat{\mathbf{y}}_{o}|\mathbf{q}_{1:L})}[R(\hat{\mathbf{y}}_{o},\mathbf{y}_{o})
-\beta \mathbb{D}_{KL}[f_{\theta}(\hat{\mathbf{y}}_{o}|\mathbf{q}_{1:L})||f_{ref}(\hat{\mathbf{y}}_{o}|\mathbf{q}_{1:L})]],$}
\label{eq:2}
\end{equation}
where $\hat{\mathbf{y}}_{o}$, $\mathbf{y}_{o}$ denote the predicted and ground-truth data $\hat{\mathbf{y}}_{L+1:L+H}$, $\mathbf{y}_{L+1:L+H}$ here for brevity, $\mathbf{q}_{1:L}=(\mathbf{y}_{1:L},\mathbf{x}_{1:L})$ denotes the historical context. $R(\cdot)$ is a time series reward function evaluating forecasting quality. The pretrained TSFM is used as the reference policy $f_{ref}(\cdot)$. $\beta$ controls the KL regularization strength. The KL constraint prevents excessive policy updates toward abnormal high-reward predictions and improves training stability.

We adopt Group Relative Policy Optimization (GRPO) \cite{shao2024deepseekmath, guo2025deepseek} to optimize the RFT objective, as it avoids introducing an additional neural value model for advantage estimation. By leveraging available ground-truth forecasting labels, GRPO enables fine-grained reward and advantage computation for individual prediction steps. Given $G$ as the number of sampled forecasts within each group, the corresponding step-wise learning objective is:
% \begin{equation}
% \centering
% \begin{aligned}
% \mathcal{L}_{RFT}(\theta)=\mathbb{E}_{\mathbf{q}_{1:L}\sim \mathcal{D},\{\hat{\mathbf{y}}_{o}^{k}\}_{k=1}^{G}\sim f_{old}(\hat{\mathbf{y}}_{o}|\mathbf{q}_{1:L})}[\frac{1}{GH}\sum_{k=1}^{G}\sum_{t=1}^{N_{p}}\min[\phi_{t}^{(k)}(\theta)A_{t}^{(k)}, \\
% \mathrm{clip}(\phi_{t}^{(k)}(\theta),1-\varepsilon,1+\varepsilon)A_{t}^{(k)}]-\beta \mathbb{D}_{KL}[f_{\theta}(\hat{\mathbf{y}}_{t}^{(k)}|\mathbf{q}_{<t})||f_{ref}(\hat{\mathbf{y}}_{t}^{(k)}|\mathbf{q}_{<t})]].
% \end{aligned}
% \label{eq:3}
% \end{equation}
\begin{equation}
\centering
\resizebox{0.44\textwidth}{!}{$
\begin{aligned}
\mathcal{L}_{RFT}(\theta)=\mathbb{E}_{\mathbf{q}_{1:L}\sim \mathcal{D},\{\hat{\mathbf{y}}_{o}^{k}\}_{k=1}^{G}\sim f_{old}(\cdot)}[\frac{1}{GH}\sum_{k=1}^{G}\sum_{t=1}^{N_{p}}\min[\phi_{t}^{(k)}(\theta)A_{t}^{(k)}, \\
\mathrm{clip}(\phi_{t}^{(k)}(\theta),1-\varepsilon,1+\varepsilon)A_{t}^{(k)}]-\beta \mathbb{D}_{KL}[f_{\theta}(\cdot)||f_{ref}(\cdot)]].
\end{aligned}
$}
\label{eq:3}
\end{equation}

Here, $t$ denotes the TSFM decoding step, where each step predicts a patch $\hat{\mathbf{y}}_{t} \in \mathbb{R}^{p\times N_{d}}$ with size $p$ rather than a single timestamp $i$, and $N_p$ is the number of predicted patches. $\phi_{t}^{(k)}(\theta)=\frac{f_{\theta}(\hat{\mathbf{y}}_{t}^{(k)}|\mathbf{q}_{<t})}{f_{old}(\hat{\mathbf{y}}_{t}^{(k)}|\mathbf{q}_{<t})}$ represents the importance sampling ratio between the updated and previous policies, and $\mathbf{q}_{<t}$ denotes the produced sequence until step $t$. $\varepsilon$ is the clipping coefficient of policy gradients. The per-step KL penalty term can be calculated by an unbiased estimator \cite{shao2024deepseekmath}: $\mathbb{D}_{KL}[f_{\theta}(\hat{\mathbf{y}}_{t}^{(k)}|\mathbf{q}_{<t})||f_{ref}(\hat{\mathbf{y}}_{t}^{(k)}|\mathbf{q}_{<t})]=\frac{f_{ref}(\cdot)}{f_{\theta}(\cdot)}-\log \frac{f_{ref}(\cdot)}{f_{\theta}(\cdot)}-1$. In next section, we will specify the fine-grained advantage computation $A_{t}^{(k)}$ should narrow down to each prediction step and variate, and integrate external ground-truth guidance, which is crucial for reliably assessing the contribution of on-policy generated forecasts and steering TSFMs towards the correct exploration direction.

Different from conventional SFT, RFT optimizes the likelihood of on-policy forecasts $\hat{\mathbf{y}}_{L+1:L+H}^{(k)}$ generated from on-the-fly $f_{\theta}(\cdot)$, which may naturally introduce temporal distribution shifts beyond the original training samples, as depicted in Figure \ref{fig:ft_paradigms}(b). Such out-of-distribution self-exploration alleviates the risk of overfitting and enables TSFMs to discover more robust forecasting behaviors, thus improving their generalization towards unseen temporal patterns.

\section{TimeRFT Training}
\label{sec:methodology}
\begin{figure*}[ht]
\centering
\includegraphics[width=0.96\linewidth]{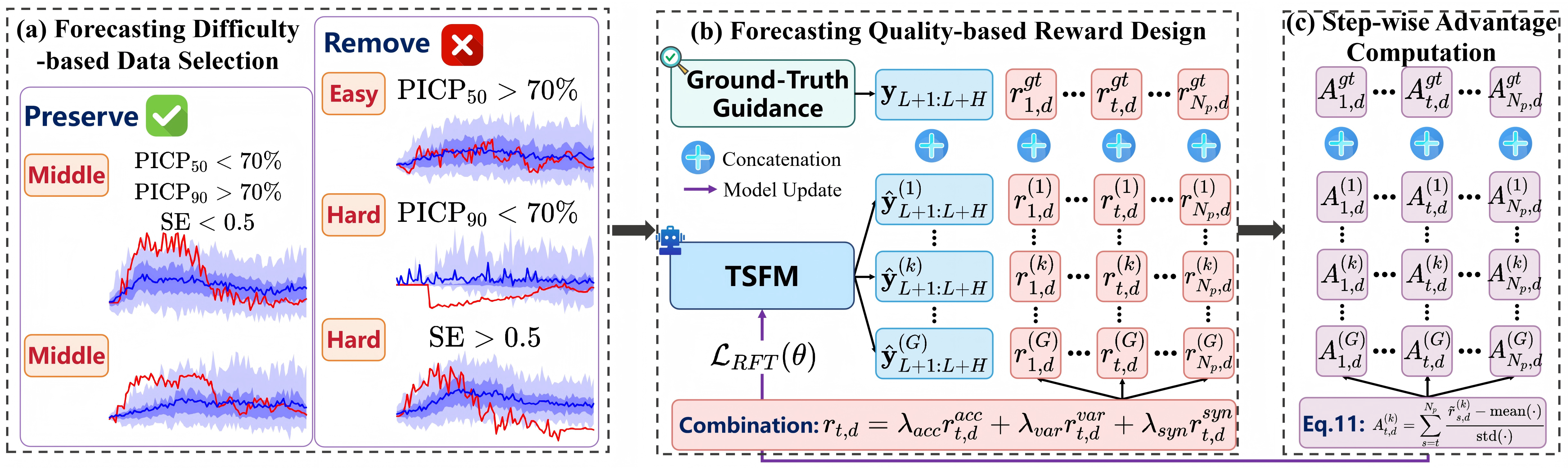}
\vspace{-6pt}
\caption{Overview of TimeRFT. It first selects informative time series samples with suitable forecasting difficulty to provide effective GRPO training signals. Then, fine-grained temporal rewards and group-normalized advantages are computed at each prediction step to guide policy optimization through the RFT objective.}
\label{fig:framework}
\end{figure*}

To enable effective time series RFT for TSFM adaptation, we propose TimeRFT, which introduces two forecasting-oriented training strategies beyond the direct application of naive GRPO. First, we develop a \textbf{forecasting quality-based step-wise temporal reward mechanism}, which provides fine-grained and multi-faceted evaluation for each prediction step and generates reliable advantage signals for on-policy trajectory optimization. Second, we design a \textbf{forecasting difficulty-based data selection strategy}, which identifies informative training samples with appropriate forecastability and filters out uninformative series that provide insufficient learning signals. The overall framework and algorithm of TimeRFT is illustrated in Figure \ref{fig:framework} and Appendix \ref{sec:algorithm} respectively.

\subsection{Forecasting Quality-based Reward Design}
\label{sec:reward_design}
Effective time series RFT requires reliable reward signals to evaluate the quality of on-policy generated forecasts $\{\hat{\mathbf{y}}_{L+1:L+H}^{(k)}\}_{k=1}^{G}$ at the granularity of each prediction step $t$ and variate $d$. Such fine-grained evaluation enables reliable advantage estimation and guides TSFMs toward generating forecasts with both numerical accuracy and realistic temporal dynamics. The key challenge is therefore to determine the contribution of each predicted patch $\hat{\mathbf{y}}_{t,d}^{(k)}$ rather than merely evaluating the accuracy of the predicted sequence as a whole. Unlike LLM reasoning tasks that often rely on sparse outcome rewards due to the difficulty of modeling intermediate reasoning processes \cite{gao2024designing, guo2025deepseek}, time series forecasting naturally provides step-wise ground-truth observations, enabling fine-grained temporal reward modeling without introducing additional neural reward models. Therefore, we design a forecasting quality-based temporal reward mechanism that evaluates each predicted patch $\hat{\mathbf{y}}_{t,d}$ from multiple complementary perspectives, including point-wise accuracy, local variability consistency and global frequency alignment. For brevity, the superscript $k$ indicating different on-policy sampled sequences is omitted in the following reward formulations.

\subsubsection{Accuracy Reward}
Point-wise accuracy is the fundamental criterion for evaluating prediction quality. We derive the accuracy reward based on normalized Mean Squared Error (nMSE):
\begin{equation}
\centering
r_{t,d}^{acc}=\exp (-\frac{1}{p} \sum_{j=1}^{p} (\hat{\mathbf{y}'}_{(t-1)p+j+L,d}-\mathbf{y}'_{(t-1)p+j+L,d})^{2}),
\label{eq:4}
\end{equation}
where $\hat{\mathbf{y}'}_{L+1:L+H,d}$ and $\mathbf{y}'_{L+1:L+H,d}$ denote normalized predicted and ground-truth sequences. The normalization based on the target sequence statistics (i.e. mean and standard deviation of $\{\mathbf{y}_{i,d}\}_{i=L+1}^{L+H}$) reduces the influence of different data magnitudes on reward distributions. The exponential transformation $\exp(-(\cdot))$ constrains $r_{t,d}^{acc}$ into $[0,1]$, assigning lower rewards to forecasts with larger prediction errors.

However, accuracy-only rewards may overlook important temporal structures and encourage over-smoothed forecasts, such as flat outputs or incorrect trends, when solely pursuing for minimizing average point-wise prediction errors \cite{kudrat2025patch, qiu2025dbloss, wang2026quadratic}. Therefore, we further introduce complementary temporal characteristic-based rewards below.

\subsubsection{Variability Reward}
\label{sec:var_reward}
To distinguish  high-quality forecasts, local variability characterizes short-term fluctuations, and it is an essential temporal property~\cite{aksu2024gift, kudrat2025patch}. In order to encourage generated forecasts to preserve local temporal structures, we leverage a dispersion-based metric \cite{kudrat2025patch} to measure the discrepancy of local variability between predicted and ground-truth patches:
\begin{equation}
\centering
r_{t,d}^{var}=\exp(-\mathbb{D}_{\mathrm{KL}}[\varphi(\hat{\mathbf{y}'}_{t,d})||\varphi(\mathbf{y}'_{t,d})]),
\label{eq:5}
\end{equation}
where $\varphi(\cdot)$ applies the softmax function to transform patch values into discrete distributions, and the KL divergence measures the variability difference between predicted $\hat{\mathbf{y}'}_{t,d}$ and actual $\mathbf{y}'_{t,d}$ patches. By optimizing patch-level variability consistency, TimeRFT encourages TSFMs to capture local temporal fluctuations beyond point-wise accuracy.

\subsubsection{Frequency Reward}
\label{sec:freq_reward}
While accuracy and variability rewards focus on point-level and local temporal dynamics, they may disregard long-range dependencies and high-frequency components \cite{wang2025fredf, tao2026memcast}. To characterize global temporal structures, we introduce a reweighted sequence-wise frequency-domain reward:
\begin{equation}
\centering
\resizebox{0.44\textwidth}{!}{$
r_{t,d}^{freq}=\exp (-\frac{1}{N_{\xi}} \sum_{\xi=1}^{N_{\xi}}w_{\xi}(\mathcal{F}(\hat{\mathbf{y}'}_{L+1:L+H,d})(\xi)-\mathcal{F}(\mathbf{y}'_{L+1:L+H,d})(\xi))^{2}),
$}
\label{eq:6}
\end{equation}
where $\mathcal{F}(\cdot)$ denotes the Fast Fourier Transform, $\xi$ denotes the frequency component with a total number $N_{\xi}$. The weight coefficient $w_{\xi}=\frac{\exp(\xi)}{\sum_{j=1}^{N_{\xi}}\exp(\xi)}$ is obtained through softmax normalization over frequencies, assigning larger weights to high-frequency components. This design encourages TSFMs to preserve global temporal correlations and fine-grained variations in the frequency domain.

\subsubsection{Synergistic Reward Modeling}
\label{sec:syn_reward}
Above reward design captures complementary aspects of forecasting quality. And instead of simply combining them in an additive fashion, we explicitly model the interaction between accuracy and temporal characteristics through a multiplicative synergy bonus:
\begin{equation}
\centering
r_{t,d}^{syn}=r_{t,d}^{acc}\times r_{t,d}^{var}+r_{t,d}^{acc}\times r_{t,d}^{freq},
\label{eq:7}
\end{equation}
where $r_{t,d}^{syn}$ reinforces forecasts that simultaneously achieve high numerical accuracy via $r_{t,d}^{acc}$ and preserve realistic local and global temporal properties via $r_{t,d}^{var}$, $r_{t,d}^{freq}$.

\subsubsection{Aggregated Reward}
Putting them together, our designed step-wise temporal reward is formulated as follows:
\begin{equation}
\centering
r_{t,d}=\lambda_{acc}r_{t,d}^{acc}+\lambda_{var}r_{t,d}^{var}+\lambda_{syn}r_{t,d}^{syn},
\label{eq:8}
\end{equation}
where $\lambda_{acc}$, $\lambda_{var}$ and $\lambda_{syn}$ are weighting coefficients. The overall reward for an entire forecast trajectory in Equation \ref{eq:2} is computed as $R(\cdot)=\sum_{d=1}^{N_{d}}\sum_{t=1}^{N_{p}}r_{t,d}$. By integrating accuracy, local variability and global frequency consistency, the designed reward mechanism offers a holistic evaluation of on-policy forecasts and enables reliable step-wise credit assignment for time series RFT of TSFMs. Since $r_{t,d}^{freq}$ is computed at the sequence level, it is incorporated into the synergy term rather than directly optimized as a dense reward, avoiding interference with fine-grained patch-level reward signals.

\subsection{Refined Advantage Estimation}
\label{sec:advantage_estimation}
Standard GRPO for LLM reasoning typically estimates advantages through group-wise reward normalization \cite{zhang2025grpo, shao2024deepseekmath}. This design works well when pretrained LLMs can already produce many correct intermediate answers and hinge on self-exploration to move toward better reasoning trajectories. However, pretrained TSFMs are generally much less reliable at generating accurate forecasts by self-evolution alone, so low-quality forecasts with small rewards may still obtain positive advantages when they dominate in group sampling. Such misaligned credit assignment can destabilize RFT training and misguide TSFMs away from effective exploration directions. To address this issue, we refine the original group-normalized scheme by explicitly incorporating ground-truth forecasts into advantage estimation, together with piecewise reward shaping to stabilize optimization. We also adopt a step-wise advantage computation that is naturally suited to the fine-grained reward design.

\subsubsection{Ground-truth Guidance Incorporation}
To help TSFMs identify the correct evolution direction during RFT, we augment each on-policy reward group with the ground-truth forecast as an external guidance:
\begin{equation}
\centering
R_{t,d}=\{r_{t,d}^{(1)},...,r_{t,d}^{(k)},...,r_{t,d}^{(G)}\}\cup \{r_{t,d}^{gt}\},
\label{eq:9}
\end{equation}
where $r_{t,d}^{gt}=r_{t,d}^{(G+1)}$ denotes the maximal reward assigned to the ground-truth labels, and $R_{t,d}$ is the extended reward group at step $t$ and variate $d$.

\subsubsection{Piecewise Reward Shaping}
\label{sec:reward_shaping}
Although ground-truth guidance provides a useful optimization target, directly using it may excessively enforce TSFMs to fit targets beyond their current capacity, leading to unstable training or mode collapse \cite{yan2025learning}. We therefore apply a piecewise reward shaping function to compress the negative effect of overly large $r_{t,d}^{gt}$:
\begin{equation}
\centering
\tilde{r}_{t,d}^{(k)}=\begin{cases}
\tau+\alpha\cdot \mathrm{In}((r_{t,d}^{(k)}-\tau)+1), & r_{t,d}^{(k)}\ge\tau \\
r_{t,d}^{(k)}, & r_{t,d}^{(k)}<\tau \end{cases}
\label{eq:10}
\end{equation}
where $\tau$ is the reward threshold and $\alpha$ is the truncation coefficient. The reshaped reward group is written as
$\tilde{R}_{t,d}=\{\tilde{r}_{t,d}^{(1)},...,\tilde{r}_{t,d}^{(G)},\tilde{r}_{t,d}^{gt}\}$.

\subsubsection{Step-wise Advantage Computation}
Following the step-wise advantage estimation method in \cite{cui2025process}, we compute fine-grained credit assignment for each on-policy forecast within its group as:
\begin{equation}
\centering
A_{t,d}^{(k)}=\sum_{s=t}^{N_{p}} \frac{\tilde{r}_{s,d}^{(k)}-\mathrm{mean}(\{\frac{\sum_{j=1}^{N_{p}}\tilde{r}_{j,d}^{(n)}}{N_{p}}\}_{n=1}^{G+1})}{\mathrm{std}(\{\frac{\sum_{j=1}^{N_{p}}\tilde{r}_{j,d}^{(n)}}{N_{p}}\}_{n=1}^{G+1})},
\label{eq:11}
\end{equation}
where the mean and standard deviation are computed across the extended reward group over $N_p$ patches. This formulation is well matched to autoregressive time series forecasting, since each prediction step should be evaluated not only by its immediate quality but also by its influence on future predictions, thereby reducing error accumulation and improving the stability of RFT training.

\subsection{Forecasting Difficulty-based Data 
Selection}
\label{sec:data_selection}
Identifying informative training samples is crucial for effective RFT on TSFMs. Real-world time series often contain noise and uncertainty induced by exogenous events \cite{fu2025selective, cheng2023weakly}, such as device outages in solar forecasting. Training on such corrupted samples can cause TSFMs to fit non-generalizable temporal patterns, especially when the amount of downstream data is limited. Besides, samples that are either too easy or too difficult to forecast provide weak learning signals for GRPO \cite{sun2025improving}, since they fail to yield discriminative and stable reward signals within each group. When a sample is already well captured by the pretrained TSFM, its rewards are close to the ground truth and further exploration brings limited benefit. Conversely, when a sample is far beyond the model's current forecasting capacity, rewards from on-policy samples become unstable and may lead to poor policy updates or mode collapse. Hence, TimeRFT favors training samples with moderate forecasting difficulty, which provide more informative and stable exploration directions.

To this end, we propose a forecasting difficulty-based data selection strategy that filters out samples with non-generalizable prediction patterns or weak learning signals. We quantify forecasting difficulty from both model-based and statistics-based perspectives, and define the corresponding data selection criteria below.

\subsubsection{Model-based Selection}
To align with the self-exploratory nature of RFT, we measure forecasting difficulty using the initial zero-shot performance of the pretrained TSFM on each training sample. In RL-based LLM post-training, task difficulty is often estimated by the group correctness ratio \cite{zhang2025grpo}, i.e., the proportion of correct responses within total $G$ responses of a group. For TSFM forecasting, we instead use Prediction Interval Coverage Probability (PICP) for ground-truth targets \cite{li2024transformer} as a proxy for forecasting difficulty, since it reflects how well the predictive distribution represented by a pretrained TSFM matches actual observations. This indicates how difficult for a TSFM to learn realistic temporal patterns. PICP can be calculated as follows:
\begin{equation}
\centering
\mathrm{PICP}=\frac{1}{N_{d}H} \sum_{d=1}^{N_{d}}\sum_{i=1}^{H}\mathbb{I}_{\mathbf{y}_{i,d}\ge\hat{\mathbf{y}}_{i,d}^{low}}\cdot\mathbb{I}_{\mathbf{y}_{i,d}\le \hat{\mathbf{y}}_{i,d}^{high}},
\label{eq:12}
\end{equation}
where $\hat{\mathbf{y}}_{i,d}^{low}$, $\hat{\mathbf{y}}_{i,d}^{high}$ denotes the point-wise lower and upper bounds of the prediction interval. We identify training samples as easy to forecast if $\mathrm{PICP}_{50}>70\%$, where $\mathrm{PICP}_{50}$ indicates $25\%$ and $75\%$ quantiles as bounds. This suggests if the relatively sharp $50\%$ interval can cover most parts of a ground-truth target, it is easy for the pretrained TSFM to generate high-quality on-policy forecasts, thus offering limited training signals. Conversely, difficult training samples can be identified by $\mathrm{PICP}_{90}<70\%$, where $\mathrm{PICP}_{90}$ indicates $5\%$ and $95\%$ quantiles as bounds. This implies that if a relatively wide $90\%$ interval fails to cover most parts of a ground-truth target, it may exceed current forecasting capacity of the pretrained TSFM and produce unstable policy gradients. The remaining samples with moderate forecasting difficulty are retained for RFT training.

\subsubsection{Statistics-based Selection}
In addition to model-based filtering, we further employ Spectral Entropy (SE) to assess the forecastability of time series data \cite{aksu2024gift}. SE characterizes the complexity of temporal patterns in the frequency domain: lower SE indicates a more predictable series with a higher signal-to-noise ratio, while higher SE suggests a more complex and less forecastable series. We filter out low-forecastability samples by requiring $\mathrm{SE}(\mathbf{y}_{1:L+H})>0.5$, where SE is normalized to $[0,1]$ in the frequency domain.

\section{Experiments}
\begin{table*}[ht]
\centering
\caption{Overall comparison between TimeRFT and baseline methods across diverse forecasting tasks and training data regimes.}
\vspace{-6pt}
\label{tab:overall}
\resizebox{1.0\textwidth}{!}{
\begin{tabular}{c|c|cccc|cccccc|cccccc}
\toprule
\multirow{3}{*}{\makecell{Data\\Size}} & \multirow{3}{*}{Methods} & \multicolumn{4}{c|}{Univariate Forecasting}                                            & \multicolumn{6}{c|}{Multivariate Forecasting}                                                                                                 & \multicolumn{6}{c}{Covariate-informed Forecasting}                                                                                             \\ \cmidrule{3-18} 
                           &                          & \multicolumn{2}{c|}{\makecell{Loop\\Seattle}}                    & \multicolumn{2}{c|}{ERCOT}      & \multicolumn{2}{c|}{ETT}                             & \multicolumn{2}{c|}{\makecell{Jena\\Weather}}                    & \multicolumn{2}{c|}{BOOMLET}    & \multicolumn{2}{c|}{\makecell{UCI Air\\Quality}}                 & \multicolumn{2}{c|}{\makecell{Solar with\\Weather}}              & \multicolumn{2}{c}{\makecell{ENTSO-e\\Load}} \\ \cmidrule{3-18} 
                           &                          & MSE            & \multicolumn{1}{c|}{MAE}            & MSE            & MAE            & MSE            & \multicolumn{1}{c|}{MAE}            & MSE            & \multicolumn{1}{c|}{MAE}            & MSE            & MAE            & MSE            & \multicolumn{1}{c|}{MAE}            & MSE            & \multicolumn{1}{c|}{MAE}            & MSE             & MAE            \\ \midrule
0\%                        & Pretrain                 & 4.186          & \multicolumn{1}{c|}{3.971}          & 2.047          & 10.896         & 10.909         & \multicolumn{1}{c|}{1.717}          & 9.891          & \multicolumn{1}{c|}{3.827}          & 2.322          & 6.722          & 2.937          & \multicolumn{1}{c|}{1.312}          & 8.647          & \multicolumn{1}{c|}{6.431}          & 22.832          & 11.876         \\ \midrule
\multirow{8}{*}{5\%}       & MSD-Mixer                & 4.973          & \multicolumn{1}{c|}{5.470}          & 4.737          & 16.899         & 12.945         & \multicolumn{1}{c|}{2.050}          & 15.254         & \multicolumn{1}{c|}{5.589}          & 2.572          & 8.589          & 7.116          & \multicolumn{1}{c|}{2.100}          & 6.619          & \multicolumn{1}{c|}{6.042}          & 60.304          & 19.559         \\
                           & iTransformer             & 5.229          & \multicolumn{1}{c|}{5.627}          & 5.059          & 17.539         & 12.346         & \multicolumn{1}{c|}{2.009}          & 15.301         & \multicolumn{1}{c|}{5.580}          & 2.570          & 8.574          & 7.065          & \multicolumn{1}{c|}{2.091}          & 6.349          & \multicolumn{1}{c|}{5.856}          & 60.765          & 19.595         \\
                           & Memformer                & 5.094          & \multicolumn{1}{c|}{5.545}          & 4.884          & 17.197         & 11.818         & \multicolumn{1}{c|}{1.970}          & 15.375         & \multicolumn{1}{c|}{5.575}          & 2.564          & 8.531          & 7.020          & \multicolumn{1}{c|}{2.082}          & 6.123          & \multicolumn{1}{c|}{5.695}          & 61.469          & 19.660         \\
                           & TimeSFT                  & 2.996          & \multicolumn{1}{c|}{3.625}          & 1.651          & 9.852          & 5.928          & \multicolumn{1}{c|}{1.249}          & 9.309          & \multicolumn{1}{c|}{3.932}          & 2.388          & 7.279          & \textbf{2.693} & \multicolumn{1}{c|}{\textbf{1.229}} & {\ul 4.625}    & \multicolumn{1}{c|}{3.842}          & 16.066          & 8.860          \\
                           & TimeLP                   & 3.186          & \multicolumn{1}{c|}{3.629}          & 1.724          & 10.072         & 6.028          & \multicolumn{1}{c|}{1.261}          & 9.928          & \multicolumn{1}{c|}{3.957}          & 2.511          & 7.479          & 2.823          & \multicolumn{1}{c|}{1.264}          & 4.854          & \multicolumn{1}{c|}{3.899}          & 17.109          & 9.231          \\
                           & TimeLoRA                 & 3.011          & \multicolumn{1}{c|}{3.622}          & 1.660          & 9.877          & 5.955          & \multicolumn{1}{c|}{1.247}          & 9.293          & \multicolumn{1}{c|}{3.937}          & 2.378          & 7.257          & {\ul 2.701}    & \multicolumn{1}{c|}{{\ul 1.232}}    & 4.695          & \multicolumn{1}{c|}{3.861}          & {\ul 15.801}    & {\ul 8.836}    \\
                           & \textbf{TimeGRPO}        & {\ul 2.921}    & \multicolumn{1}{c|}{{\ul 3.622}}    & {\ul 1.617}    & {\ul 9.842}    & {\ul 5.907}    & \multicolumn{1}{c|}{{\ul 1.234}}    & {\ul 8.502}    & \multicolumn{1}{c|}{{\ul 3.787}}    & {\ul 2.322}    & {\ul 7.080}    & 2.793          & \multicolumn{1}{c|}{1.258}          & 4.675          & \multicolumn{1}{c|}{{\ul 3.722}}    & 16.104          & 8.906          \\
                           & \textbf{TimeRFT}         & \textbf{2.757} & \multicolumn{1}{c|}{\textbf{3.400}} & \textbf{1.571} & \textbf{9.690} & \textbf{5.550} & \multicolumn{1}{c|}{\textbf{1.187}} & \textbf{8.394} & \multicolumn{1}{c|}{\textbf{3.730}} & \textbf{2.166} & \textbf{6.470} & 2.771          & \multicolumn{1}{c|}{1.254}          & \textbf{4.528} & \multicolumn{1}{c|}{\textbf{3.680}} & \textbf{15.329} & \textbf{8.643} \\ \midrule
\multirow{8}{*}{20\%}      & MSD-Mixer                & 4.230          & \multicolumn{1}{c|}{4.974}          & 3.032          & 13.201         & 7.904          & \multicolumn{1}{c|}{1.653}          & 14.270         & \multicolumn{1}{c|}{5.157}          & 2.473          & 8.218          & 6.562          & \multicolumn{1}{c|}{2.013}          & 4.758          & \multicolumn{1}{c|}{4.614}          & 30.150          & 13.018         \\
                           & iTransformer             & 4.420          & \multicolumn{1}{c|}{5.107}          & 3.424          & 14.087         & 7.498          & \multicolumn{1}{c|}{1.608}          & 14.259         & \multicolumn{1}{c|}{5.127}          & 2.410          & 7.997          & 6.461          & \multicolumn{1}{c|}{1.996}          & 4.639          & \multicolumn{1}{c|}{4.479}          & 23.319          & 11.517         \\
                           & Memformer                & 4.317          & \multicolumn{1}{c|}{5.036}          & 3.195          & 13.575         & 7.180          & \multicolumn{1}{c|}{1.569}          & 14.188         & \multicolumn{1}{c|}{5.087}          & 2.407          & 7.946          & 6.369          & \multicolumn{1}{c|}{1.981}          & 4.540          & \multicolumn{1}{c|}{4.355}          & 26.218          & 12.328         \\
                           & TimeSFT                  & 2.614          & \multicolumn{1}{c|}{3.553}          & 1.529          & 9.526          & 5.683          & \multicolumn{1}{c|}{1.195}          & 10.253         & \multicolumn{1}{c|}{4.076}          & 2.309          & 6.853          & 2.692          & \multicolumn{1}{c|}{1.226}          & 4.196          & \multicolumn{1}{c|}{3.669}          & 6.465           & 5.561          \\
                           & TimeLP                   & 2.808          & \multicolumn{1}{c|}{3.602}          & 1.584          & 9.687          & 5.724          & \multicolumn{1}{c|}{1.200}          & 10.966         & \multicolumn{1}{c|}{4.068}          & 2.318          & 6.876          & 2.707          & \multicolumn{1}{c|}{1.236}          & 4.615          & \multicolumn{1}{c|}{3.818}          & 7.216           & 5.760          \\
                           & TimeLoRA                 & 2.697          & \multicolumn{1}{c|}{3.584}          & 1.539          & 9.571          & 5.696          & \multicolumn{1}{c|}{1.195}          & 10.338         & \multicolumn{1}{c|}{4.079}          & 2.308          & 6.866          & 2.689          & \multicolumn{1}{c|}{1.225}          & {\ul 4.173}    & \multicolumn{1}{c|}{3.662}          & {\ul 6.291}     & {\ul 5.498}    \\
                           & \textbf{TimeGRPO}        & {\ul 2.440}    & \multicolumn{1}{c|}{{\ul 3.432}}    & {\ul 1.503}    & {\ul 9.459}    & {\ul 5.551}    & \multicolumn{1}{c|}{{\ul 1.182}}    & {\ul 8.964}    & \multicolumn{1}{c|}{{\ul 3.887}}    & {\ul 2.256}    & {\ul 6.744}    & {\ul 2.635}    & \multicolumn{1}{c|}{{\ul 1.217}}    & 4.180          & \multicolumn{1}{c|}{{\ul 3.537}}    & 6.542           & 5.645          \\
                           & \textbf{TimeRFT}         & \textbf{2.193} & \multicolumn{1}{c|}{\textbf{3.091}} & \textbf{1.473} & \textbf{9.361} & \textbf{5.184} & \multicolumn{1}{c|}{\textbf{1.152}} & \textbf{8.610} & \multicolumn{1}{c|}{\textbf{3.824}} & \textbf{2.145} & \textbf{6.160} & \textbf{2.593} & \multicolumn{1}{c|}{\textbf{1.195}} & \textbf{4.132} & \multicolumn{1}{c|}{\textbf{3.496}} & \textbf{4.196}  & \textbf{4.525} \\ \midrule
\multirow{8}{*}{50\%}      & MSD-Mixer                & 3.359          & \multicolumn{1}{c|}{4.291}          & 1.897          & 10.625         & 5.968          & \multicolumn{1}{c|}{1.432}          & 12.292         & \multicolumn{1}{c|}{4.564}          & 2.471          & 8.196          & 5.453          & \multicolumn{1}{c|}{1.835}          & 4.169          & \multicolumn{1}{c|}{3.956}          & 20.804          & 11.020         \\
                           & iTransformer             & 3.557          & \multicolumn{1}{c|}{4.450}          & 2.089          & 11.037         & 5.624          & \multicolumn{1}{c|}{1.383}          & 12.360         & \multicolumn{1}{c|}{4.554}          & 2.358          & 7.801          & 5.318          & \multicolumn{1}{c|}{1.810}          & 4.118          & \multicolumn{1}{c|}{3.936}          & 22.258          & 11.510         \\
                           & Memformer                & 3.456          & \multicolumn{1}{c|}{4.369}          & 2.047          & 11.064         & 5.503          & \multicolumn{1}{c|}{1.363}          & 12.198         & \multicolumn{1}{c|}{4.515}          & 2.314          & 7.637          & 5.201          & \multicolumn{1}{c|}{1.789}          & 4.099          & \multicolumn{1}{c|}{3.912}          & 16.200          & 9.793          \\
                           & TimeSFT                  & 2.535          & \multicolumn{1}{c|}{3.379}          & 1.543          & 9.599          & 5.466          & \multicolumn{1}{c|}{1.197}          & 9.931          & \multicolumn{1}{c|}{3.931}          & 2.389          & 6.738          & 2.607          & \multicolumn{1}{c|}{1.205}          & 4.357          & \multicolumn{1}{c|}{3.596}          & 6.682           & 5.800          \\
                           & TimeLP                   & 2.764          & \multicolumn{1}{c|}{3.617}          & 1.569          & 9.645          & 5.504          & \multicolumn{1}{c|}{1.202}          & 10.284         & \multicolumn{1}{c|}{3.930}          & 2.501          & 6.726          & 2.781          & \multicolumn{1}{c|}{1.220}          & 4.605          & \multicolumn{1}{c|}{3.646}          & 7.035           & 5.761          \\
                           & TimeLoRA                 & 2.565          & \multicolumn{1}{c|}{3.370}          & 1.548          & 9.586          & {\ul 5.465}    & \multicolumn{1}{c|}{1.192}          & 9.895          & \multicolumn{1}{c|}{3.940}          & 2.380          & 6.726          & 2.610          & \multicolumn{1}{c|}{1.202}          & 4.379          & \multicolumn{1}{c|}{3.612}          & 6.770           & 5.869          \\
                           & \textbf{TimeGRPO}        & {\ul 2.350}    & \multicolumn{1}{c|}{{\ul 3.323}}    & {\ul 1.539}    & {\ul 9.558}    & 5.596          & \multicolumn{1}{c|}{{\ul 1.161}}    & {\ul 9.480}    & \multicolumn{1}{c|}{{\ul 3.912}}    & {\ul 2.344}    & {\ul 6.671}    & {\ul 2.573}    & \multicolumn{1}{c|}{{\ul 1.201}}    & {\ul 4.150}    & \multicolumn{1}{c|}{{\ul 3.554}}    & {\ul 4.305}     & {\ul 4.616}    \\
                           & \textbf{TimeRFT}         & \textbf{2.082} & \multicolumn{1}{c|}{\textbf{3.019}} & \textbf{1.449} & \textbf{9.205} & \textbf{5.197} & \multicolumn{1}{c|}{\textbf{1.133}} & \textbf{8.910} & \multicolumn{1}{c|}{\textbf{3.837}} & \textbf{2.194} & \textbf{6.148} & \textbf{2.459} & \multicolumn{1}{c|}{\textbf{1.156}} & \textbf{3.855} & \multicolumn{1}{c|}{\textbf{3.323}} & \textbf{3.867}  & \textbf{4.189} \\ \midrule
\multirow{8}{*}{100\%}     & MSD-Mixer                & 2.922          & \multicolumn{1}{c|}{3.872}          & 1.630          & 10.090         & 5.784          & \multicolumn{1}{c|}{1.412}          & 11.739         & \multicolumn{1}{c|}{4.441}          & 2.516          & 6.666          & 4.681          & \multicolumn{1}{c|}{1.690}          & 3.918          & \multicolumn{1}{c|}{3.851}          & 19.223          & 10.184         \\
                           & iTransformer             & 3.098          & \multicolumn{1}{c|}{3.947}          & 1.668          & 10.041         & 5.397          & \multicolumn{1}{c|}{1.348}          & 11.202         & \multicolumn{1}{c|}{4.360}          & 2.400          & 6.855          & 4.599          & \multicolumn{1}{c|}{1.675}          & 3.868          & \multicolumn{1}{c|}{3.804}          & 20.564          & 10.855         \\
                           & Memformer                & 3.071          & \multicolumn{1}{c|}{4.018}          & 1.632          & 9.982          & 5.328          & \multicolumn{1}{c|}{1.328}          & 10.859         & \multicolumn{1}{c|}{4.322}          & 2.352          & 6.819          & 4.517          & \multicolumn{1}{c|}{1.659}          & 3.716          & \multicolumn{1}{c|}{3.740}          & 18.789          & 10.243         \\
                           & TimeSFT                  & 2.421          & \multicolumn{1}{c|}{3.448}          & 1.580          & 9.761          & 5.411          & \multicolumn{1}{c|}{1.206}          & 9.158          & \multicolumn{1}{c|}{3.915}          & 2.303          & 6.623          & 2.508          & \multicolumn{1}{c|}{{\ul 1.177}}    & 4.208          & \multicolumn{1}{c|}{3.544}          & 5.379           & 5.232          \\
                           & TimeLP                   & 2.540          & \multicolumn{1}{c|}{3.481}          & 1.644          & 10.013         & 5.951          & \multicolumn{1}{c|}{1.235}          & 9.559          & \multicolumn{1}{c|}{3.922}          & 2.514          & 6.717          & 2.600          & \multicolumn{1}{c|}{1.203}          & 4.449          & \multicolumn{1}{c|}{3.629}          & 5.765           & 5.446          \\
                           & TimeLoRA                 & 2.434          & \multicolumn{1}{c|}{3.464}          & 1.562          & {\ul 9.721}    & 5.492          & \multicolumn{1}{c|}{1.206}          & {\ul 9.101}    & \multicolumn{1}{c|}{{\ul 3.901}}    & 2.317          & 6.797          & 2.518          & \multicolumn{1}{c|}{1.180}          & 4.250          & \multicolumn{1}{c|}{3.556}          & 5.348           & 5.209          \\
                           & \textbf{TimeGRPO}        & {\ul 2.276}    & \multicolumn{1}{c|}{{\ul 3.303}}    & {\ul 1.555}    & 9.760          & {\ul 5.269}    & \multicolumn{1}{c|}{{\ul 1.161}}    & 9.222          & \multicolumn{1}{c|}{3.927}          & {\ul 2.261}    & {\ul 6.128}    & {\ul 2.504}    & \multicolumn{1}{c|}{1.184}          & {\ul 4.129}    & \multicolumn{1}{c|}{{\ul 3.412}}    & {\ul 4.105}     & {\ul 4.554}    \\
                           & \textbf{TimeRFT}         & \textbf{2.032} & \multicolumn{1}{c|}{\textbf{3.021}} & \textbf{1.463} & \textbf{9.429} & \textbf{4.900} & \multicolumn{1}{c|}{\textbf{1.130}} & \textbf{8.996} & \multicolumn{1}{c|}{\textbf{3.865}} & \textbf{2.129} & \textbf{5.985} & \textbf{2.452} & \multicolumn{1}{c|}{\textbf{1.169}} & \textbf{3.862} & \multicolumn{1}{c|}{\textbf{3.322}} & \textbf{3.767}  & \textbf{4.230} \\ 
\bottomrule
\end{tabular}
}
\end{table*}

\subsection{Experimental Settings}
\subsubsection{Datasets and Forecasting Tasks}
We evaluate TimeRFT on eight real-world time series datasets from fev-bench \cite{shchur2025fev}, covering three representative forecasting tasks: univariate forecasting ($N_{d}=1, N_{c}=0$), multivariate forecasting ($N_{d}>1, N_{c}=0$) and covariate-informed forecasting ($N_{d}>0, N_{c}>0$). Following the real-world protocol in \cite{shchur2025fev}, we utilize the last $W$ non-overlapping windows of length $H$ for validation and testing, and reserve the remaining sequences for training. This setting preserves realistic temporal structure and reflects practical deployment scenarios with varying forecast horizons and data availability. Dataset statistics are summarized in Table \ref{tab:dataset}.

\subsubsection{Baseline Methods}
We compare TimeRFT with three groups of baselines. First, we include three non-pretrained forecasting models that jointly model temporal and cross-channel dependencies: MSD-Mixer \cite{zhong2024multi}, iTransformer \cite{liu2024itransformer} and Memformer \cite{cheng2024memory}. Second, we compare against SFT-based TSFM adaptation methods, including full parameter-based TimeSFT \cite{na2026timepeft}, head-only TimeLP \cite{goswami2024moment} and LoRA-based TimeLoRA \cite{gupta2024beyond}. Third, we include TimeGRPO as the naive RFT baseline, which removes the reward design and data selection from TimeRFT while keeping the same GRPO method.

\subsubsection{Evaluation Metrics}
For each test instance, we sample 100 forecasts from the learned predictive distribution and use their mean as the point prediction $\hat{\mathbf{y}}_{L+1:L+H}$. Following the evaluation principle in \cite{fan2023dish}, we report Mean Squared Error (MSE) and Mean Absolute Error (MAE) metrics in the raw data value space rather than in normalized space. This can directly reflect distribution shifts of raw time point values, avoiding the impact of data normalization. The dataset-specific scaling factors are provided in Table \ref{tab:metric_scale}.

\subsubsection{TSFM backbones} 
Our primary TSFM backbone is MOIRAI-MoE \cite{liu2025moirai-moe}, which supports both autoregressive forecasting and probabilistic sampling, making it well suited for RFT. Unless otherwise stated, we use MOIRAI-MoE$_{\mathrm{S}}$ for the main experiments to balance performance and efficiency. Besides, we validate TimeRFT on decoder-only ToTo-1.0 \cite{cohen2026time} and encoder-only Chronos-2 \cite{ansari2025chronos} to demonstrate TimeRFT is compatible with other TSFM architectures.

\subsubsection{Implementation Details}
The context length is set to $L=mH$ with $m\in[2,20]$ akin to \cite{liu2025moirai-moe}. The KL strength and group size are set to $\beta=0.001$ and $G=8$. Following vanilla GRPO \cite{shao2024deepseekmath}, we set $f_{old}(\cdot)=f_{\theta}(\cdot)$, which makes the importance ratio equal to one and removes the need for gradient clipping. The reward weights are empirically set to $\lambda_{acc}=0.9$, $\lambda_{var}=0.1$, and $\lambda_{syn}=0.01$; the reward shaping parameters are set to $\alpha=0.01$ and $\tau=0.8$. We optimize with Adam using a learning rate of $5e^{-6}$, weight decay of $0.1$ and batch size 128. All experiments are run on a single NVIDIA RTX PRO 6000 GPU with 96GB memory.

\subsection{Main Results}
\subsubsection{Overall Comparison} 
As shown in Table \ref{tab:overall}, RFT-based methods consistently outperform SFT-based methods in most forecasting settings, and TimeRFT yields average improvements of 10.17\% in MSE and 7.04\% in MAE versus TimeSFT and TimeLoRA. These significant gains indicate proposed reward-driven self-exploration of RFT induces stronger generalizability than direct ground-truth imitation of SFT under temporal distribution shifts. The only notable exception is UCI Air Quality under the 5\% few-shot regime, where both RFT-based methods struggle due to the limited exploration space and weak sample diversity in extremely low-data settings.

\subsubsection{Compatibility to Other TSFMs}
we further compare TimeRFT with another adapter-based finetuning method called Time-PEFT \cite{na2026timepeft} on Chronos-2 and ToTo-1.0. As shown in Table \ref{tab:compatible_short}, TimeRFT remains consistently superior to both TimeSFT and Time-PEFT across all three datasets under both limited-data and full-data regimes, indicating that the proposed RFT training recipes are architecture-agnostic rather than tailored to a specific TSFM. Refer to Table \ref{tab:compatible_full}, \ref{tab:overall_base} in Appendix \ref{sec:add_main_results} for more comparison results on different backbones.
\begin{table}[ht]
\centering
\caption{Compatibility to other TSFM backbones.}
\vspace{-6pt}
\label{tab:compatible_short}
\resizebox{0.478\textwidth}{!}{
\begin{tabular}{cccccccc}
\toprule
\multirow{2}{*}{Backbones} & \multirow{2}{*}{Methods} & \multicolumn{2}{c}{Loop Seattle} & \multicolumn{2}{c}{ETT}         & \multicolumn{2}{c}{ENTSO-e Load} \\ \cmidrule{3-8} 
                           &                          & MSE             & MAE            & MSE            & MAE            & MSE             & MAE            \\ \midrule
\multirow{3}{*}{Chronos-2} & TimeSFT                  & 2.545           & 3.544          & 5.658          & 1.219          & 5.992           & 5.415          \\
                           & Time-PEFT                 & 2.487           & 3.435          & 5.553          & 1.213          & 5.946           & 5.393          \\
                           & TimeRFT                  & \textbf{2.136}  & \textbf{3.047} & \textbf{5.037} & \textbf{1.146} & \textbf{4.150}  & \textbf{4.523} \\ \midrule
\multirow{3}{*}{ToTo-1.0}  & TimeSFT                  & 2.545           & 3.465          & 5.652          & 1.217          & 5.908           & 5.398          \\
                           & Time-PEFT                 & 2.480           & 3.435          & 5.551          & 1.213          & 5.755           & 5.273          \\
                           & TimeRFT                  & \textbf{2.120}  & \textbf{3.067} & \textbf{5.013} & \textbf{1.144} & \textbf{4.079}  & \textbf{4.433} \\ 
\bottomrule
\end{tabular}
}
\end{table}

\subsection{Zero-shot Transferability Study}
\label{sec:transfer_study}
We further evaluate whether the proposed finetuning methods can transfer learned temporal patterns from one dataset to another related but unseen domain without additional adaptation. Such zero-shot cross-data transfer is practically valuable when target-domain observations are limited or unavailable \cite{chen2024similarity, qiu2025dbloss}. we construct three realistic transfer settings with inherent domain correlations: Loop Seattle across different traffic sensors (s1$\rightarrow$s2), ETT across different power transformer stations (m1$\rightarrow$m2), and ENTSO-e Load across different geographical regions (r1$\rightarrow$r2). The consistent gains of TimeRFT shown in Table \ref{tab:transfer_short} indicate that the proposed forecasting-oriented RFT training learns more domain-generalizable prediction patterns, rather than overfitting to source-domain temporal modes. Refer to Table \ref{tab:transfer_full} in Appendix \ref{sec:add_transfer_results} for more results and analysis on this zero-shot transfer study.
\begin{table}[ht]
\centering
\caption{Zero-shot transferability to unseen datasets.}
\vspace{-6pt}
\label{tab:transfer_short}
\resizebox{0.428\textwidth}{!}{
\begin{tabular}{ccccccc}
\toprule
\multirow{2}{*}{Methods} & \multicolumn{2}{c}{\makecell{Loop Seattle\\(s1->s2)}} & \multicolumn{2}{c}{\makecell{ETT\\(m1->m2)}} & \multicolumn{2}{c}{\makecell{ENTSO-e Load\\(r1->r2)}} \\ \cmidrule{2-7} 
                         & MSE                       & MAE                       & MSE                   & MAE                  & MSE                       & MAE                       \\ \midrule
Pretrain                 & 9.179                     & 5.762                     & 9.792                 & 2.061                & 2.696                     & 12.558                    \\
TimeSFT                  & 5.070                     & 4.740                     & 7.805                 & 1.839                & 1.795                     & {\ul 9.870}               \\
TimeLP                   & 5.554                     & 4.965                     & 8.248                 & 1.896                & 1.972                     & 10.237                    \\
TimeLoRA                 & 5.083                     & 4.767                     & 7.789                 & 1.837                & {\ul 1.789}               & 9.909                     \\
TimeGRPO                 & {\ul 4.884}               & {\ul 4.537}               & {\ul 7.676}           & {\ul 1.810}          & 1.804                     & 9.871                     \\
TimeRFT                  & \textbf{4.641}            & \textbf{4.297}            & \textbf{7.566}        & \textbf{1.793}       & \textbf{1.656}            & \textbf{9.553}            \\ 
\bottomrule
\end{tabular}
}
\vspace{-10pt}
\end{table}

\subsection{Ablation Study}
Table \ref{tab:ablation_short} shows that every component of TimeRFT contributes to the final performance. First, the variability reward is beneficial for capturing local temporal structure. Replacing it with (denoted as "rw/") patch-wise correlation or mean-value rewards \cite{kudrat2025patch} degrades accuracy, indicating that simple patch statistics are insufficient to model useful local dynamics. Second, the sequence-wise frequency reward is more effective than its patch-wise counterpart, confirming that global spectral alignment provides stronger supervision than local frequency matching. Third, removing the synergy term causes a noticeable drop on all datasets, showing that combining accuracy with temporal-structure rewards is more effective than optimizing them independently. Besides, data selection is the most critical design as removing it leads to the largest performance degradation, which validates the importance of filtering uninformative samples before RFT training. Lastly, reward shaping improves training stability and final accuracy by preventing overly large rewards from dominating optimization, while KL penalty provides an additional gain by constraining excessive policy drift. Refer to Table \ref{tab:ablation_full}, \ref{tab:sft_full} in Appendix \ref{sec:add_ablation_results} for more validations on key designs in TimeRFT.
\begin{table}[ht]
\centering
\caption{Ablation results on key designs of TimeRFT.}
\vspace{-6pt}
\label{tab:ablation_short}
\resizebox{0.478\textwidth}{!}{
\begin{tabular}{ccccccc}
\toprule
\multirow{2}{*}{Methods}  & \multicolumn{2}{c}{Loop Seattle} & \multicolumn{2}{c}{ETT}         & \multicolumn{2}{c}{ENTSO-e Load} \\ \cmidrule{2-7} 
                          & MSE             & MAE            & MSE            & MAE            & MSE             & MAE            \\ \midrule
w/o variability $r^{var}_{t,d}$                & 2.182           & 3.189          & 5.246          & 1.146          & 4.385           & 4.581          \\
rw/ correlation $r^{corr}_{t,d}$               & 2.139           & 3.079          & 5.137          & \textbf{1.140} & 4.079           & 4.433          \\
rw/ mean $r^{mean}_{t,d}$               & 2.181           & 3.128          & 5.278          & 1.146          & 4.242           & 4.568          \\ \midrule
w/o sequence-wise $r^{freq}_{t,d}$ & 2.160           & 3.109          & 5.151          & 1.146          & 4.218           & 4.527          \\
rw/ patch-wise $r^{freq}_{t,d}$    & 2.142           & 3.078          & 5.055          & 1.143          & 4.173           & 4.471          \\ \midrule
w/o $r^{syn}_{t,d}$                & 2.179           & 3.184          & 5.213          & 1.149          & 4.349           & 4.571          \\ \midrule
w/o data selection        & 2.300           & 3.249          & 5.333          & 1.160          & 4.949           & 4.916          \\ \midrule
w/o reward shaping        & 2.209           & 3.216          & 5.319          & 1.150          & 4.562           & 4.715          \\ \midrule
w/o KL constraint         & 2.157           & 3.123          & 5.161          & 1.142          & 4.225           & 4.544          \\ \midrule
TimeRFT                   & \textbf{2.113}  & \textbf{3.056} & \textbf{5.042} & 1.141          & \textbf{3.982}  & \textbf{4.378} \\ 
\bottomrule
\end{tabular}
}
\end{table}

\begin{figure*}[ht]
\centering
\includegraphics[width=0.98\linewidth]{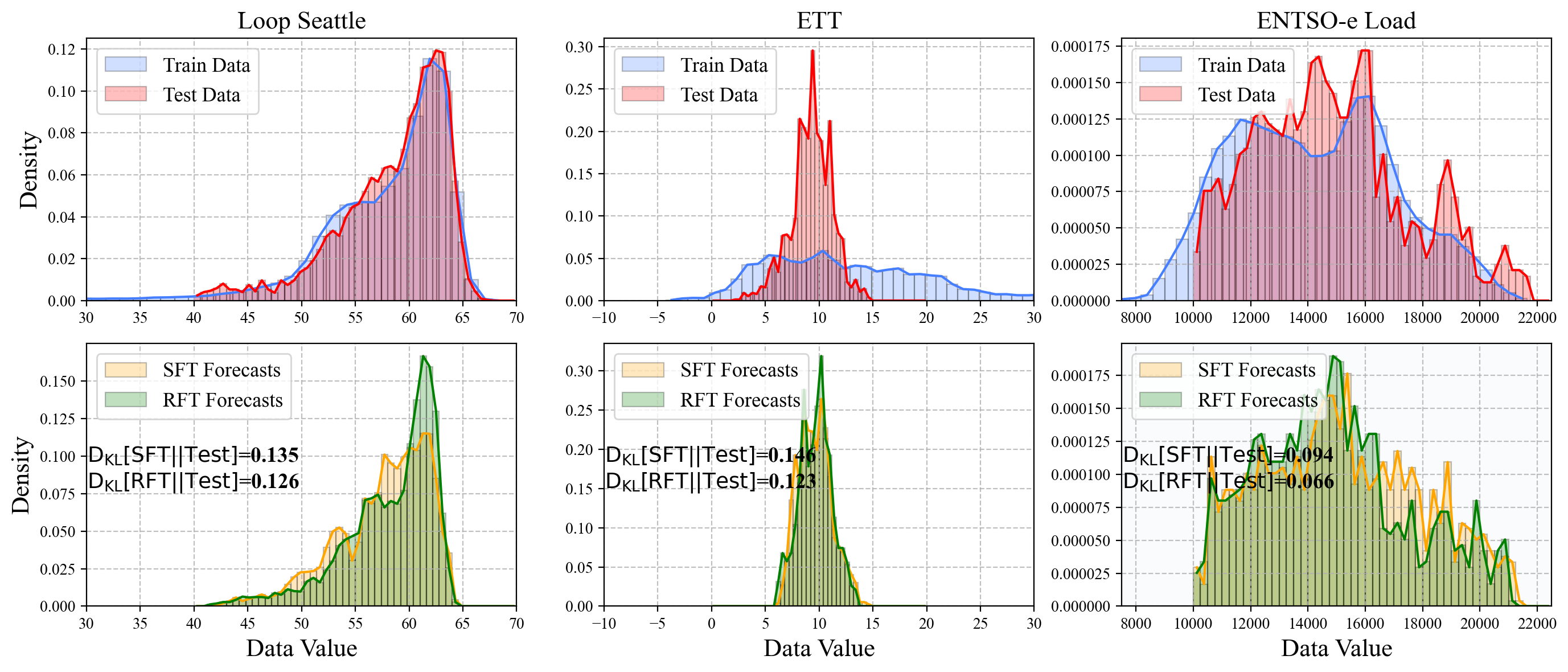}
\vspace{-15pt}
\caption{Upper: distribution mismatch between historical training and future test data. Lower: TimeRFT produces forecasts that better match the real test distribution.}
\vspace{-12pt}
\label{fig:distribution_shifts_raw}
\end{figure*}

\subsection{Method Analysis}
Figure \ref{fig:distribution_shifts_raw} reveals that even within the same dataset, the future test distribution is not identical to the historical training distribution, indicating the presence of temporal distribution shifts. Although the shifts are moderate in these in-domain settings, the learned predictive distribution from TimeSFT still tends to remain closer to the training pattern, whereas TimeRFT consistently aligns better with the test distribution. This is reflected by TimeRFT's lower KL divergence towards the real test distribution on all three datasets. This provides direct evidence that TimeRFT better handles temporal distribution shifts rather than simply fitting the observed training data. Refer to Figure \ref{fig:distribution_shifts_transfer} in Appendix \ref{sec:add_distribution_results} for more distribution shift results pertaining to the zero-shot transfer study in Section \ref{sec:transfer_study}, and Appendix \ref{sec:add_method_analysis} for more analysis of the TimeRFT method.

\section{Conclusion}
In this work, we propose TimeRFT, a new reinforcement finetuning method for adapting TSFMs to downstream forecasting tasks. By leveraging on-policy exploration and forecasting-oriented training strategies, TimeRFT effectively alleviates the overfitting issue of conventional SFT-based adaptation and improves forecasting generalization under temporal distribution shifts and diverse data regimes. Specifically, the proposed forecasting quality-based reward design and forecasting difficulty-based data selection strategy provide informative learning signals and efficient exploration guidance, enabling TimeRFT to achieve consistent performance improvements across various forecasting scenarios.

In future work, we will investigate more efficient TimeRFT optimization strategies by reducing the training cost through sparse TSFM parameter updates \cite{mukherjee2025reinforcement} and more efficient on-policy sampling strategies that generate fewer but more informative forecasting trajectories \cite{xu2025not}. Furthermore, we plan to extend TimeRFT beyond forecasting tasks toward broader time series understanding and reasoning scenarios, such as multimodal time series analysis \cite{xie2025chatts}.

\bibliographystyle{ACM-Reference-Format}
\bibliography{reference}

% \clearpage
\appendix
\section{TimeRFT Algorithm}
\label{sec:algorithm}
Algorithm \ref{algo:timerft} details the procedural implementation of the proposed TimeRFT training method, corresponding to the visual framework illustrated
in Figure \ref{fig:framework} and two forecasting-oriented training strategies clarified in Section \ref{sec:methodology}.
\begin{algorithm}
\small
\caption{TimeRFT Training Method.}
\label{algo:timerft}
\textbf{Input:} Training set $\mathcal{D}$, pretrained TSFM $f_{\theta}(\cdot)$.\\
\textbf{Output:} TSFM adapted by TimeRFT.\\
\nl \Repeat{traverse $\mathcal{D}$}{
  \nl Sample $\{\mathbf{x}_{1:L},\mathbf{y}_{1:L+H}\}$ from $\mathcal{D}$.\\
  \nl Infer $\{\mathbf{x}_{1:L},\mathbf{y}_{1:L+H}\}$ by initial $f_{\theta}(\cdot)$.\\
  \nl \uIf{$\mathrm{PICP}_{50}>70\%$ $\mathrm{or}$ $\mathrm{PICP}_{90}<70\%$ $\mathrm{or}$ $\mathrm{SE}>0.5$}{
  \nl Remove $\{\mathbf{x}_{1:L},\mathbf{y}_{1:L+H}\}$ from $\mathcal{D}$.\\
  }
}
\nl \Repeat{convergence}{
  \nl Sample batch $\mathcal{D}_{b}$ from $\mathcal{D}$.\\
  \nl Generate on-policy group $\{\hat{\mathbf{y}}_{L+1:L+H}^{(k)}\}_{k=1}^{G}$ for $\mathbf{y}_{1:L}$ in $\mathcal{D}_{b}$.\\
  \nl Compute step-wise reward $r_{t,d}^{(k)}$ using Equation \ref{eq:8}.\\
  \nl Reshape step-wise reward $\tilde{r}_{t,d}^{(k)}$ using Equation \ref{eq:10}.\\
  \nl Compute step-wise advantage $A_{t,d}^{(k)}$ using Equation \ref{eq:11}.\\
  \nl Compute RFT loss $\mathcal{L}_{RFT}(\theta)$ using Equation \ref{eq:3}.\\
  \nl Back propagate policy gradients.
}
\nl \Return{Finetuned TSFM.}
\end{algorithm}

\section{Additional Experimental Settings}
\label{sec:add_exp_settings}
We cover various fields of real-world time series datasets to compare different TSFM finetuning methods. The dataset statistics and construction are specified in Table \ref{tab:dataset}. Besides, to explicitly present temporal distribution shifts in raw data value space, we report the scaling factors of MSE/MAE metrics across different datasets are reported in Table \ref{tab:metric_scale}. An exceptional case is that the MSE metric of ENTSO-e Load dataset in Table \ref{tab:transfer_short}, \ref{tab:transfer_full} scaled by $1e^{-6}$.
\begin{table}[ht]
\centering
\caption{Real-world dataset statistics and usage.}
\vspace{-6pt}
\label{tab:dataset}
\resizebox{0.478\textwidth}{!}{
\begin{tabular}{cccccccc}
\toprule
Dataset            & \multicolumn{1}{l}{Domain} & Rate  & $T$    & $W$ & $H$ & $N_{d}$ & $N_{c}$ \\ \midrule
Loop Seattle       & Mobility                   & 5min  & 93600  & 20  & 288 & 1       & 0       \\
ERCOT              & Energy                     & 1h    & 148152 & 20  & 168 & 1       & 0       \\
ETT                & Energy                     & 15min & 65840  & 20  & 96  & 7       & 0       \\
Jena Weather       & Nature                     & 10min & 46944  & 20  & 144 & 21      & 0       \\
BOOMLET-963        & Cloud                      & 1min  & 13984  & 20  & 60  & 28      & 0       \\
UCI Air Quality    & Nature                     & 1h    & 7917   & 10  & 72  & 4       & 3       \\
Solar with Weather & Energy                     & 15min & 69192  & 20  & 96  & 1       & 9       \\
ENTSO-e Load       & Energy                     & 30min & 85725  & 20  & 48  & 1       & 3       \\ 
\bottomrule
\end{tabular}
}
\end{table}
\begin{table}[ht]
\centering
\caption{The scaling factors of the magnitude of two evaluation metrics across different datasets.}
\vspace{-6pt}
\label{tab:metric_scale}
\resizebox{0.478\textwidth}{!}{
\begin{tabular}{c|cccccccc}
\toprule
Dataset & \multicolumn{2}{c}{\makecell{Loop\\Seattle}} & \multicolumn{2}{c}{ERCOT}           & \multicolumn{2}{c}{ETT}                & \multicolumn{2}{c}{\makecell{Jena\\Weather}} \\ \midrule
Metric  & MSE             & MAE            & MSE              & MAE              & MSE                & MAE               & MSE             & MAE            \\ \midrule
Scale   & $1e^{-1}$       & $1e^{0}$       & $1e^{-6}$        & $1e^{-2}$        & $1e^{0}$           & $1e^{0}$          & $1e^{-1}$       & $1e^{0}$       \\ \midrule
Dataset & \multicolumn{2}{c}{BOOMLET}      & \multicolumn{2}{c}{\makecell{UCI Air\\Quality}} & \multicolumn{2}{c}{\makecell{Solar with\\Weather}} & \multicolumn{2}{c}{\makecell{ENTSO-e\\Load}} \\ \midrule
Metric  & MSE             & MAE            & MSE              & MAE              & MSE                & MAE               & MSE             & MAE            \\ \midrule
Scale   & $1e^{0}$        & $1e^{1}$       & $1e^{-4}$        & $1e^{-2}$        & $1e^{-5}$          & $1e^{-2}$         & $1e^{-5}$       & $1e^{-2}$      \\ 
\bottomrule
\end{tabular}
}
\end{table}

\section{Additional Main Results}
\label{sec:add_main_results}
Table \ref{tab:compatible_full} reports the compatibility evaluation of TimeRFT on Chronos-2 and ToTo-1.0 on 20\% and 100\% data regimes. The main observation is consistent across both backbones: TimeRFT outperforms TimeSFT and Time-PEFT on all three datasets, confirming that the proposed TimeRFT training strategies transfers beyond MOIRAI-MoE and generalizes across different TSFM architectures. We further extend the comparison to the larger-scale MOIRAI-MoE$_{\mathrm{B}}$ under 5\% and 20\% few-shot settings across three datasets. As shown in Table \ref{tab:overall_base}, TimeRFT consistently achieves the best forecasting performance, reducing MSE/MAE by an average of 8.83\%/3.81\% over the second-best method and 15.61\%/5.37\% over the top two SFT-based methods.
\begin{table}[ht]
\centering
\caption{Full results of TSFM compatibility study in Table \ref{tab:compatible_short}, which is averaged over two data regimes.}
\vspace{-6pt}
\label{tab:compatible_full}
\resizebox{0.478\textwidth}{!}{
\begin{tabular}{ccccccccc}
\toprule
\multirow{2}{*}{Backbones} & \multirow{2}{*}{Methods}  & \multirow{2}{*}{\makecell{Data\\Size}} & \multicolumn{2}{c}{\makecell{Loop\\Seattle}} & \multicolumn{2}{c}{ETT}         & \multicolumn{2}{c}{\makecell{ENTSO-e\\Load}} \\ \cmidrule{4-9} 
                           &                           &                            & MSE             & MAE            & MSE            & MAE            & MSE             & MAE            \\ \midrule
\multirow{6}{*}{Chronos-2} & \multirow{2}{*}{TimeSFT}  & 20\%                       & 2.607           & 3.565          & 5.815          & 1.229          & 6.631           & 5.632          \\
                           &                           & 100\%                      & 2.483           & 3.524          & 5.501          & 1.209          & 5.353           & 5.198          \\
                           & \multirow{2}{*}{Time-PEFT} & 20\%                       & 2.562           & 3.438          & 5.732          & 1.225          & 6.581           & 5.584          \\
                           &                           & 100\%                      & 2.413           & 3.433          & 5.374          & 1.200          & 5.310           & 5.201          \\
                           & \multirow{2}{*}{TimeRFT}  & 20\%                       & \textbf{2.228}  & \textbf{3.095} & \textbf{5.241} & \textbf{1.158} & \textbf{4.394}  & \textbf{4.739} \\
                           &                           & 100\%                      & \textbf{2.045}  & \textbf{2.999} & \textbf{4.833} & \textbf{1.135} & \textbf{3.906}  & \textbf{4.308} \\ \midrule
\multirow{6}{*}{ToTo-1.0}  & \multirow{2}{*}{TimeSFT}  & 20\%                       & 2.605           & 3.469          & 5.811          & 1.227          & 6.642           & 5.641          \\
                           &                           & 100\%                      & 2.485           & 3.460          & 5.492          & 1.206          & 5.175           & 5.155          \\
                           & \multirow{2}{*}{Time-PEFT} & 20\%                       & 2.555           & 3.448          & 5.734          & 1.226          & 6.586           & 5.573          \\
                           &                           & 100\%                      & 2.405           & 3.421          & 5.369          & 1.201          & 4.923           & 4.973          \\
                           & \multirow{2}{*}{TimeRFT}  & 20\%                       & \textbf{2.199}  & \textbf{3.097} & \textbf{5.209} & \textbf{1.155} & \textbf{4.254}  & \textbf{4.509} \\
                           &                           & 100\%                      & \textbf{2.040}  & \textbf{3.036} & \textbf{4.817} & \textbf{1.133} & \textbf{3.905}  & \textbf{4.358} \\ 
\bottomrule
\end{tabular}
}
\end{table}
\begin{table}[ht]
\centering
\caption{Overall comparison between RFT-based and SFT-based adaptation methods on $\mathrm{\mathbf{MOIRAI-MoE_{B}}}$ \cite{liu2025moirai-moe} over three datasets under two few-shot settings.}
\vspace{-6pt}
\label{tab:overall_base}
\resizebox{0.478\textwidth}{!}{
\begin{tabular}{c|c|cc|cc|cc}
\toprule
\multirow{2}{*}{\makecell{Data\\Size}} & \multirow{2}{*}{Methods} & \multicolumn{2}{c|}{\makecell{Loop\\Seattle}} & \multicolumn{2}{c|}{ETT}        & \multicolumn{2}{c}{\makecell{ENTSO-e\\Load}} \\ \cmidrule{3-8} 
                           &                          & MSE             & MAE             & MSE            & MAE            & MSE             & MAE            \\ \midrule
0\%                        & Pretrain                 & 3.365           & 3.745           & 21.215         & 2.287          & 21.505          & 11.180         \\ \midrule
\multirow{5}{*}{5\%}       & TimeSFT                  & 2.305           & {\ul 3.218}     & 5.616          & {\ul 1.150}    & 12.747          & {\ul 7.624}    \\
                           & TimeLP                   & 2.920           & 3.431           & 6.055          & 1.318          & 14.633          & 8.296          \\
                           & TimeLoRA                 & 2.365           & 3.248           & 5.685          & 1.151          & 12.977          & 7.679          \\
                           & TimeGRPO                 & {\ul 2.103}     & 3.311           & {\ul 5.266}    & 1.154          & {\ul 11.928}    & 7.649          \\
                           & TimeRFT                  & \textbf{1.966}  & \textbf{3.110}  & \textbf{5.033} & \textbf{1.116} & \textbf{10.902} & \textbf{7.336} \\ \midrule
\multirow{5}{*}{20\%}      & TimeSFT                  & 2.179           & 2.992           & 5.869          & 1.170          & 5.280           & 4.823          \\
                           & TimeLP                   & 2.258           & 3.058           & 6.432          & 1.357          & 5.894           & 5.154          \\
                           & TimeLoRA                 & 2.197           & 3.007           & 5.988          & {\ul 1.167}    & 5.092           & 4.850          \\
                           & TimeGRPO                 & {\ul 2.079}     & {\ul 2.931}     & {\ul 5.832}    & 1.168          & {\ul 4.441}     & {\ul 4.508}    \\
                           & TimeRFT                  & \textbf{1.852}  & \textbf{2.896}  & \textbf{5.322} & \textbf{1.126} & \textbf{3.828}  & \textbf{4.147} \\ 
\bottomrule
\end{tabular}
}
\end{table}

\section{Additional Transfer Study Results}
\label{sec:add_transfer_results}
To further investigate the generalization capability under cross-domain distribution shifts, we construct zero-shot transfer tasks where the model is trained on a source dataset and directly evaluated on a related but unseen target dataset without additional finetuning. These settings are designed to reflect practical forecasting scenarios where temporal data from new locations or systems may differ from historical observations. This zero-shot setting isolates the transferability of the finetuned TSFM and avoids confounding effects from target-domain supervision. 

Loop Seattle contains traffic sensor measurements collected from multiple roadway locations in Seattle. We use the time series from sensor 1 as the source domain and sensor 2 as the target domain. Although both sensors monitor traffic flow in the same metropolitan area, they exhibit different traffic patterns due to distinct road structures, commuting behaviors, and local traffic dynamics. The ETT dataset consists of electricity transformer temperature and related load measurements collected from multiple transformer stations. We transfer models trained on station m1 to station m2 without target-domain adaptation. Different transformer stations experience heterogeneous operating conditions and temporal variations, resulting in distribution shifts in electricity-related forecasting patterns. The ENTSO-e Load dataset contains regional electricity consumption measurements from different European regions. We train models on region 1 and directly evaluate them on region 2. Different regions exhibit distinct electricity consumption behaviors caused by geographical characteristics, population distributions and regional energy usage patterns. Overall, these three transfer scenarios cover different levels of distribution shifts, ranging from spatially nearby sensors to geographically separated regions. They provide complementary evaluations of whether finetuning methods learn domain-specific patterns or more generalizable temporal forecasting representations.

As shown in Table \ref{tab:transfer_full}, TimeRFT consistently achieves the best cross-dataset performance across all transfer pairs and source-data regimes, indicating that the proposed RFT training recipes help TSFMs learn more domain-transferable temporal dynamics. Compared with the second-best method, TimeRFT reduces MSE by 4.56\% and MAE by 3.05\% on average. Relative to TimeSFT and TimeLoRA, it further yields an average improvement of 6.41\% in MSE and 5.16\% in MAE. These gains suggest that reward-driven self-exploration is more effective than supervised adaptation for capturing predictive patterns that remain invariant across related domains.
\begin{table}[ht]
\centering
\caption{Full results of zero-shot cross-dataset transfer study shown in Table \ref{tab:transfer_short}, which is averaged across four data regimes.}
\vspace{-6pt}
\label{tab:transfer_full}
\resizebox{0.478\textwidth}{!}{
\begin{tabular}{cccccccc}
\toprule
\multirow{2}{*}{\makecell{Data\\Size}} & \multirow{2}{*}{Methods} & \multicolumn{2}{c}{\makecell{Loop Seattle\\(s1->s2)}} & \multicolumn{2}{c}{\makecell{ETT\\(m1->m2)}}         & \multicolumn{2}{c}{\makecell{ENTSO-e Load\\(r1->r2)}} \\ \cmidrule{3-8} 
                           &                          & MSE             & MAE            & MSE            & MAE            & MSE            & MAE             \\ \midrule
0\%                        & Pretrain                 & 9.179           & 5.762          & 9.792          & 2.061          & 2.696          & 12.558          \\ \midrule
\multirow{5}{*}{5\%}       & TimeSFT                  & 5.117           & 4.797          & 7.897          & 1.841          & 2.565          & 11.946          \\
                           & TimeLP                   & 5.727           & 4.787          & 8.614          & 1.916          & 2.622          & 12.079          \\
                           & TimeLoRA                 & 5.139           & 4.851          & 7.892          & 1.840          & {\ul 2.555}    & {\ul 11.888}    \\
                           & TimeGRPO                 & {\ul 5.017}     & {\ul 4.605}    & {\ul 7.881}    & {\ul 1.828}    & 2.569          & 11.913          \\
                           & TimeRFT                  & \textbf{4.791}  & \textbf{4.493} & \textbf{7.691} & \textbf{1.812} & \textbf{2.432} & \textbf{11.614} \\ \midrule
\multirow{5}{*}{20\%}      & TimeSFT                  & 4.907           & 4.667          & 7.663          & 1.819          & 1.526          & 9.267           \\
                           & TimeLP                   & 5.554           & 5.199          & 8.268          & 1.909          & 1.788          & 9.617           \\
                           & TimeLoRA                 & 4.910           & 4.693          & 7.639          & 1.816          & {\ul 1.502}    & 9.241           \\
                           & TimeGRPO                 & {\ul 4.804}     & {\ul 4.467}    & {\ul 7.551}    & {\ul 1.804}    & 1.555          & {\ul 9.171}     \\
                           & TimeRFT                  & \textbf{4.790}  & \textbf{4.339} & \textbf{7.487} & \textbf{1.791} & \textbf{1.461} & \textbf{8.965}  \\ \midrule
\multirow{5}{*}{50\%}      & TimeSFT                  & 5.220           & 4.905          & 7.988          & 1.870          & 1.470          & {\ul 9.066}     \\
                           & TimeLP                   & 5.485           & 5.115          & 8.288          & 1.915          & 1.683          & 9.582           \\
                           & TimeLoRA                 & 5.267           & 4.933          & 7.957          & 1.865          & {\ul 1.454}    & 9.067           \\
                           & TimeGRPO                 & {\ul 4.724}     & {\ul 4.533}    & {\ul 7.688}    & {\ul 1.808}    & 1.478          & 9.196           \\
                           & TimeRFT                  & \textbf{4.290}  & \textbf{4.176} & \textbf{7.630} & \textbf{1.796} & \textbf{1.372} & \textbf{8.870}  \\ \midrule
\multirow{5}{*}{100\%}     & TimeSFT                  & 5.035           & 4.592          & 7.670          & 1.829          & 1.620          & {\ul 9.202}     \\
                           & TimeLP                   & 5.450           & 4.759          & 7.823          & 1.845          & 1.793          & 9.671           \\
                           & TimeLoRA                 & 5.017           & 4.590          & 7.669          & 1.826          & 1.642          & 9.441           \\
                           & TimeGRPO                 & {\ul 4.993}     & {\ul 4.544}    & {\ul 7.583}    & {\ul 1.802}    & {\ul 1.614}    & 9.204           \\
                           & TimeRFT                  & \textbf{4.693}  & \textbf{4.181} & \textbf{7.456} & \textbf{1.772} & \textbf{1.359} & \textbf{8.764}  \\ 
\bottomrule
\end{tabular}
}
\end{table}

\begin{figure*}[ht]
\centering
\includegraphics[width=0.98\linewidth]{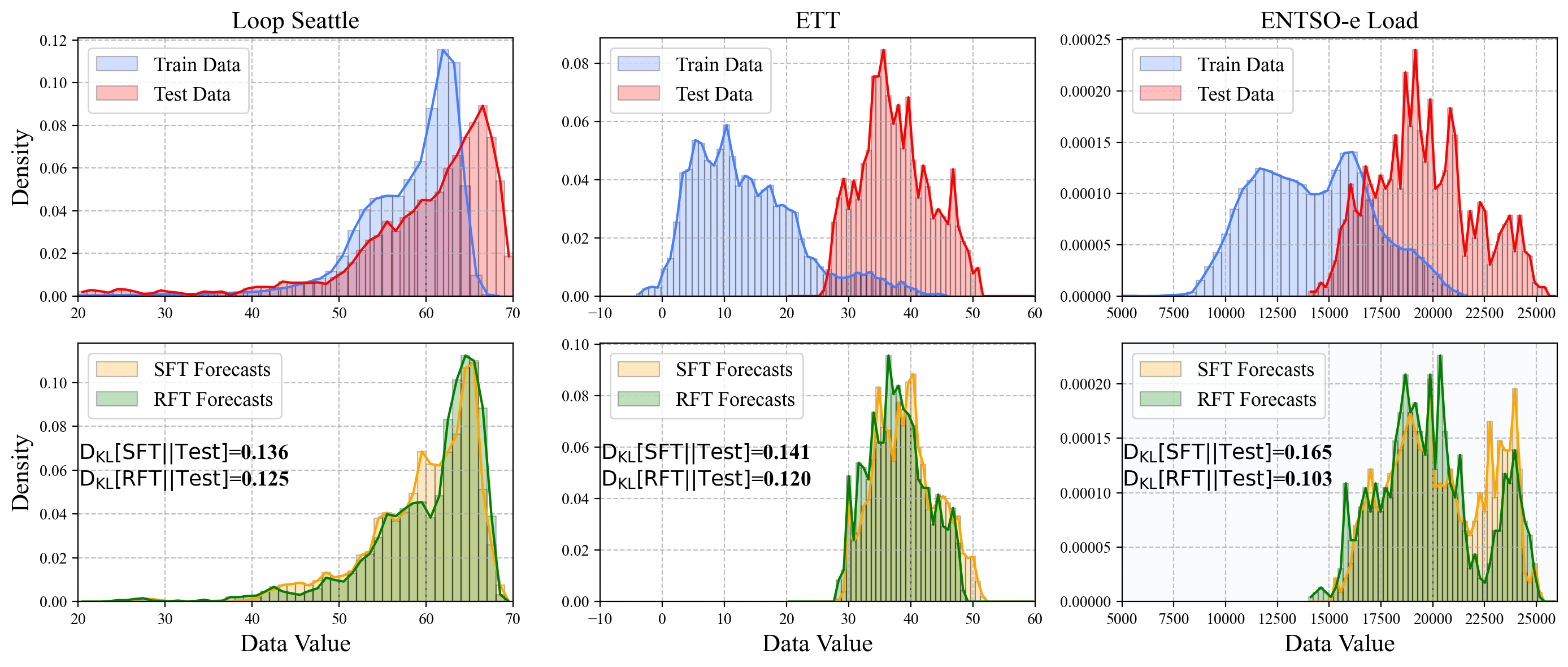}
\vspace{-6pt}
\caption{Temporal distribution shifts under the cross-dataset transfer study. The upper row compares the source training distribution and target test distribution across datasets, while the lower row shows that TimeRFT better adapts its forecast distribution to the shifted target domain than TimeSFT. The KL divergences further quantify the improved distribution alignment achieved by TimeRFT.}
\label{fig:distribution_shifts_transfer}
\end{figure*}

\section{Additional Ablation Study Results}
\label{sec:add_ablation_results}
\begin{table}[ht]
\centering
\caption{Full results of ablation study in Table \ref{tab:ablation_short}, which is averaged over two data regimes.}
\vspace{-6pt}
\label{tab:ablation_full}
\resizebox{0.478\textwidth}{!}{
\begin{tabular}{cccccccc}
\toprule
\multirow{2}{*}{Methods}                   & \multirow{2}{*}{\makecell{Data\\Size}} & \multicolumn{2}{c}{\makecell{Loop\\Seattle}} & \multicolumn{2}{c}{ETT}         & \multicolumn{2}{c}{\makecell{ENTSO-e\\Load}} \\ \cmidrule{3-8} 
                                           &                            & MSE             & MAE            & MSE            & MAE            & MSE             & MAE            \\ \midrule
\multirow{2}{*}{w/o variability $r^{var}_{t,d}$}                & 20\%                       & 2.260           & 3.253          & 5.311          & 1.160          & 4.885           & 4.838          \\
                                           & 100\%                      & 2.105           & 3.124          & 5.182          & 1.132          & 3.885           & 4.323          \\
\multirow{2}{*}{rw/ correlation $r^{corr}_{t,d}$}               & 20\%                       & 2.203           & 3.101          & 5.206          & 1.153          & 4.254           & \textbf{4.509} \\
                                           & 100\%                      & 2.075           & 3.058          & 5.068          & \textbf{1.126} & 3.905           & 4.358          \\
\multirow{2}{*}{rw/ mean $r^{mean}_{t,d}$}               & 20\%                       & 2.224           & 3.160          & 5.313          & 1.160          & 4.291           & 4.600          \\
                                           & 100\%                      & 2.139           & 3.096          & 5.244          & 1.132          & 4.192           & 4.537          \\ \midrule
\multirow{2}{*}{w/o sequence-wise $r^{freq}_{t,d}$} & 20\%                       & 2.238           & 3.171          & 5.215          & 1.156          & 4.623           & 4.689          \\
                                           & 100\%                      & 2.083           & 3.047          & 5.087          & 1.136          & 3.814           & 4.365          \\
\multirow{2}{*}{rw/ patch-wise $r^{freq}_{t,d}$}    & 20\%                       & 2.209           & 3.113          & 5.203          & 1.155          & 4.549           & 4.600          \\
                                           & 100\%                      & 2.074           & 3.043          & 4.907          & 1.130          & 3.797           & 4.342          \\ \midrule
\multirow{2}{*}{w/o $r^{syn}_{t,d}$}                & 20\%                       & 2.256           & 3.248          & 5.265          & 1.159          & 4.844           & 4.795          \\
                                           & 100\%                      & 2.102           & 3.119          & 5.162          & 1.138          & 3.853           & 4.347          \\ \midrule
\multirow{2}{*}{w/o data selection}        & 20\%                       & 2.345           & 3.322          & 5.478          & 1.177          & 5.951           & 5.389          \\
                                           & 100\%                      & 2.255           & 3.176          & 5.188          & 1.143          & 3.947           & 4.442          \\ \midrule
\multirow{2}{*}{w/o reward shaping}        & 20\%                       & 2.296           & 3.290          & 5.411          & 1.161          & 5.217           & 5.022          \\
                                           & 100\%                      & 2.121           & 3.142          & 5.226          & 1.139          & 3.907           & 4.407          \\ \midrule
\multirow{2}{*}{w/o KL constraint}         & 20\%                       & 2.232           & 3.185          & 5.203          & 1.153          & 4.609           & 4.776          \\
                                           & 100\%                      & 2.081           & 3.061          & 5.120          & 1.130          & 3.842           & 4.313          \\ \midrule
\multirow{2}{*}{TimeRFT}                   & 20\%                       & \textbf{2.193}  & \textbf{3.091} & \textbf{5.184} & \textbf{1.152} & \textbf{4.196}  & 4.525          \\
                                           & 100\%                      & \textbf{2.032}  & \textbf{3.021} & \textbf{4.900} & 1.130          & \textbf{3.767}  & \textbf{4.230} \\ 
\bottomrule
\end{tabular}
}
\end{table}
\begin{table}[ht]
\centering
\caption{Transferring the proposed two training recipes in Section \ref{sec:methodology} to TimeSFT.}
\label{tab:sft_full}
\vspace{-6pt}
\resizebox{0.478\textwidth}{!}{
\begin{tabular}{cccccccc}
\toprule
\multirow{2}{*}{Methods}           & \multirow{2}{*}{\makecell{Data\\Size}} & \multicolumn{2}{c}{\makecell{Loop\\Seattle}} & \multicolumn{2}{c}{ETT} & \multicolumn{2}{c}{\makecell{ENTSO-e\\Load}} \\ \cmidrule{3-8} 
                                   &                            & MSE             & MAE            & MSE        & MAE        & MSE             & MAE            \\ \midrule
\multirow{2}{*}{TimeSFT}           & 20\%                       & 2.614           & 3.553          & 5.683      & 1.195      & 6.465           & 5.561          \\
                                   & 100\%                      & 2.421           & 3.448          & 5.411      & 1.206      & 5.379           & 5.232          \\ \midrule
\multirow{2}{*}{+ variability loss}          & 20\%                       & 2.548           & 3.532          & 5.655      & 1.194      & 6.438           & 5.496          \\
                                   & 100\%                      & 2.440           & 3.307          & 5.405      & 1.204      & 5.353           & 5.147          \\
\multirow{2}{*}{+ frequency loss}         & 20\%                       & 2.553           & 3.522          & 5.644      & 1.193      & 6.412           & 5.501          \\
                                   & 100\%                      & 2.449           & 3.310          & 5.410      & 1.205      & 5.366           & 5.135          \\ \midrule
\multirow{2}{*}{w/ data selection} & 20\%                       & 2.547           & 3.516          & 5.619      & 1.191      & 6.345           & 5.452          \\
                                   & 100\%                      & 2.373           & 3.294          & 5.395      & 1.201      & 5.244           & 5.137          \\ \midrule
\multirow{2}{*}{w/ two training recipes}    & 20\%                       & 2.513           & 3.484          & 5.583      & 1.189      & 6.267           & 5.400          \\
                                   & 100\%                      & 2.263           & 3.290          & 5.339      & 1.198      & 5.160           & 4.940          \\ 
\bottomrule
\end{tabular}
}
\end{table}
Table \ref{tab:ablation_full} shows the full results of ablation study on key modular designs for TimeRFT. To further verify the effectiveness of variability rewards on capturing local temporal structures, we replace it with other two structural metrics from \cite{kudrat2025patch}, including the Pearson correlation coefficient and mean value of each patch. The derived correlation reward $r^{corr}_{t,d}$ and mean-value reward $r^{mean}_{t,d}$ can be calculated as
$r_{t,d}^{corr}=1-\frac{{\textstyle \sum_{i=(t-1)p}^{tp}}(\hat{\mathbf{y}'}_{i,d}-\hat{\mu}_{t,d})(\mathbf{y}'_{i,d}-\mu_{t,d})}{\hat{\sigma}_{t,d}\sigma_{t,d}}$ and $r_{t,d}^{mean}=|\hat{\mu}_{t,d}-\mu_{t,d}|$ respectively, where $\hat{\mu}_{t,d}$, $\mu_{t,d}$ and $\hat{\sigma}_{t,d}$, $\sigma_{t,d}$ denote the mean and standard deviation of the predicted and ground-truth patch.

Table \ref{tab:sft_full} shows that the two proposed training recipes are not exclusive to TimeRFT. When transferred to TimeSFT, they still yield consistent improvements over vanilla TimeSFT on all three datasets and two data regimes. In particular, both the reward-inspired losses and the data-selection strategy improve forecasting accuracy, which indicates that the proposed temporal reward design and difficulty-aware filtering indeed capture useful forecasting priors rather than acting as RL-specific heuristics. Among the single components, data selection is the most effective and stable, suggesting that removing low-forecastability samples contributes materially to performance, while the variability- and frequency-based losses provide additional but smaller gains. Importantly, combining the two recipes gives the best TimeSFT variant in every setting. This supports two points simultaneously: (i) the performance gain of TimeRFT is not merely due to data selection alone, since the same recipes also help TimeSFT. (ii) RFT remains necessary to fully exploit these recipes, as evidenced by the stronger gains of the full TimeRFT method in the main experiments.

\begin{figure*}[ht]
\centering
\includegraphics[width=0.96\linewidth]{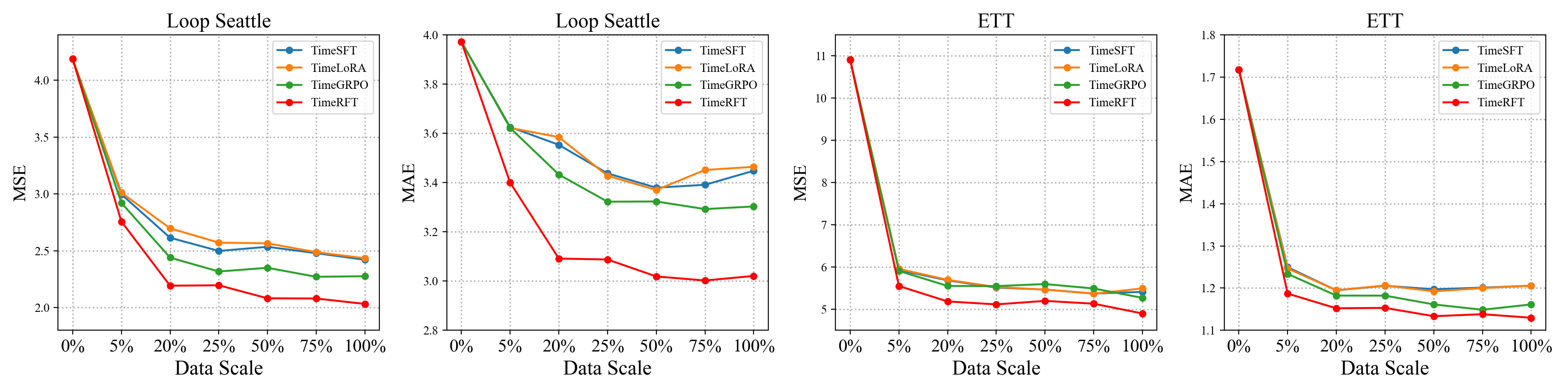}
\vspace{-10pt}
\caption{Comparison of scaling behaviors w.r.t training time series data.}
\label{fig:data_scalability}
\end{figure*}

\section{Additional Method Analysis}
\label{sec:add_method_analysis}
\subsection{Additional Distribution Shift Results}
\label{sec:add_distribution_results}
Figure \ref{fig:distribution_shifts_transfer} provides stronger evidence for the generalization effect of TimeRFT under cross-dataset transfer, where the source and target distributions exhibit larger mismatches in both location and shape. In this more challenging setting, TimeSFT remains more strongly biased toward the source-domain training distribution, while TimeRFT generates forecasts that are visibly closer to the target test distribution. The KL divergence toward the ground-truth test distribution is reduced on all three datasets: Loop Seattle ($0.136 \rightarrow 0.125$), ETT ($0.141 \rightarrow 0.120$), and ENTSO-e Load ($0.165 \rightarrow 0.103$). The larger reduction on ENTSO-e Load is particularly notable, suggesting that TimeRFT is more robust when the domain shift is substantial. Together with Figure \ref{fig:distribution_shifts_raw} in main text, these results indicate that the gains of TimeRFT are tied to improved distribution alignment under both moderate and severe temporal shifts.

\begin{figure}[ht]
\centering
\includegraphics[width=1.0\linewidth]{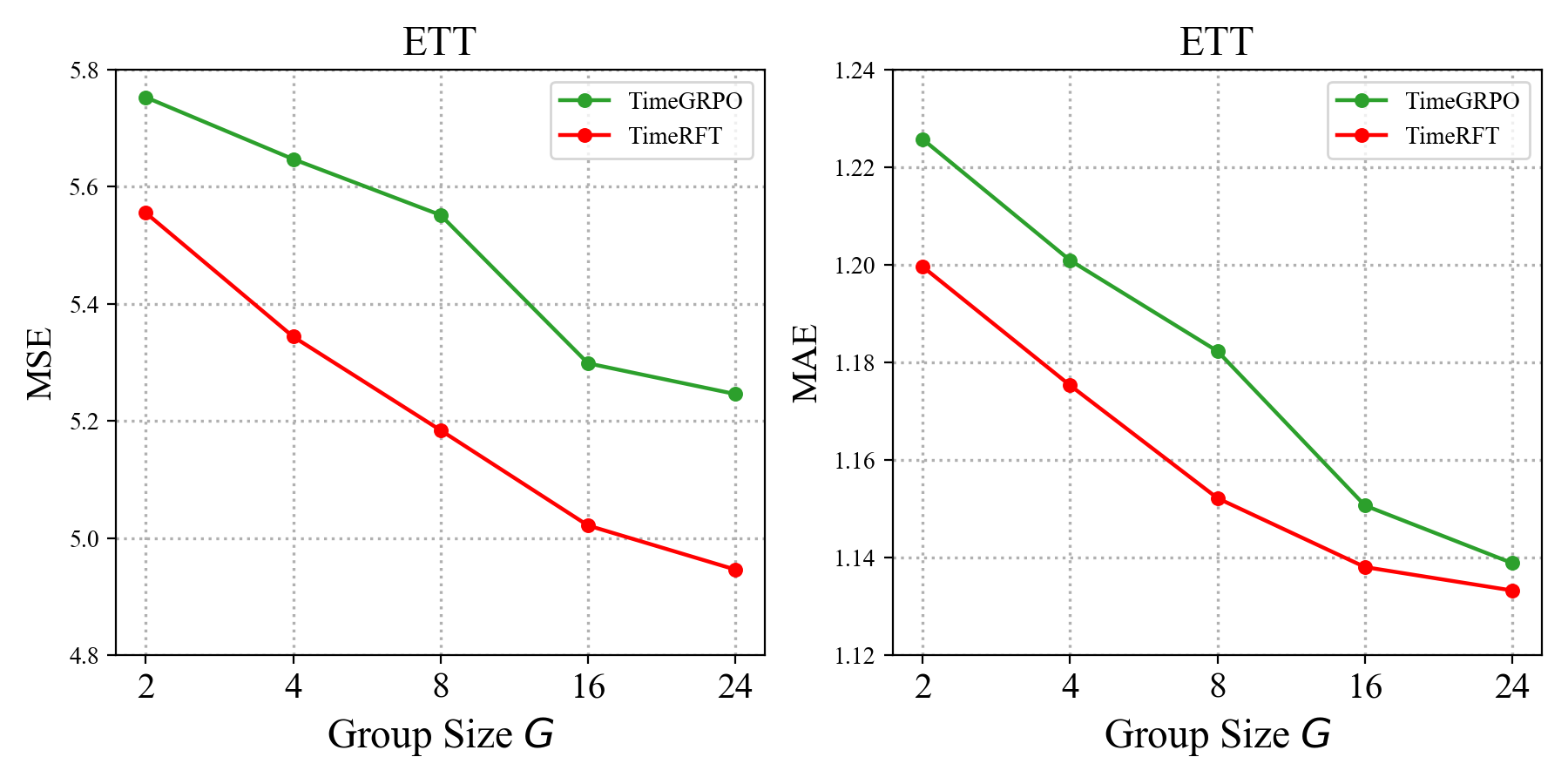}
\vspace{-10pt}
\caption{Sensitivity of RFT-based methods to group size $G$.}
\label{fig:group_size_scalability}
\end{figure}

\begin{figure}[ht]
\centering
\includegraphics[width=1.0\linewidth]{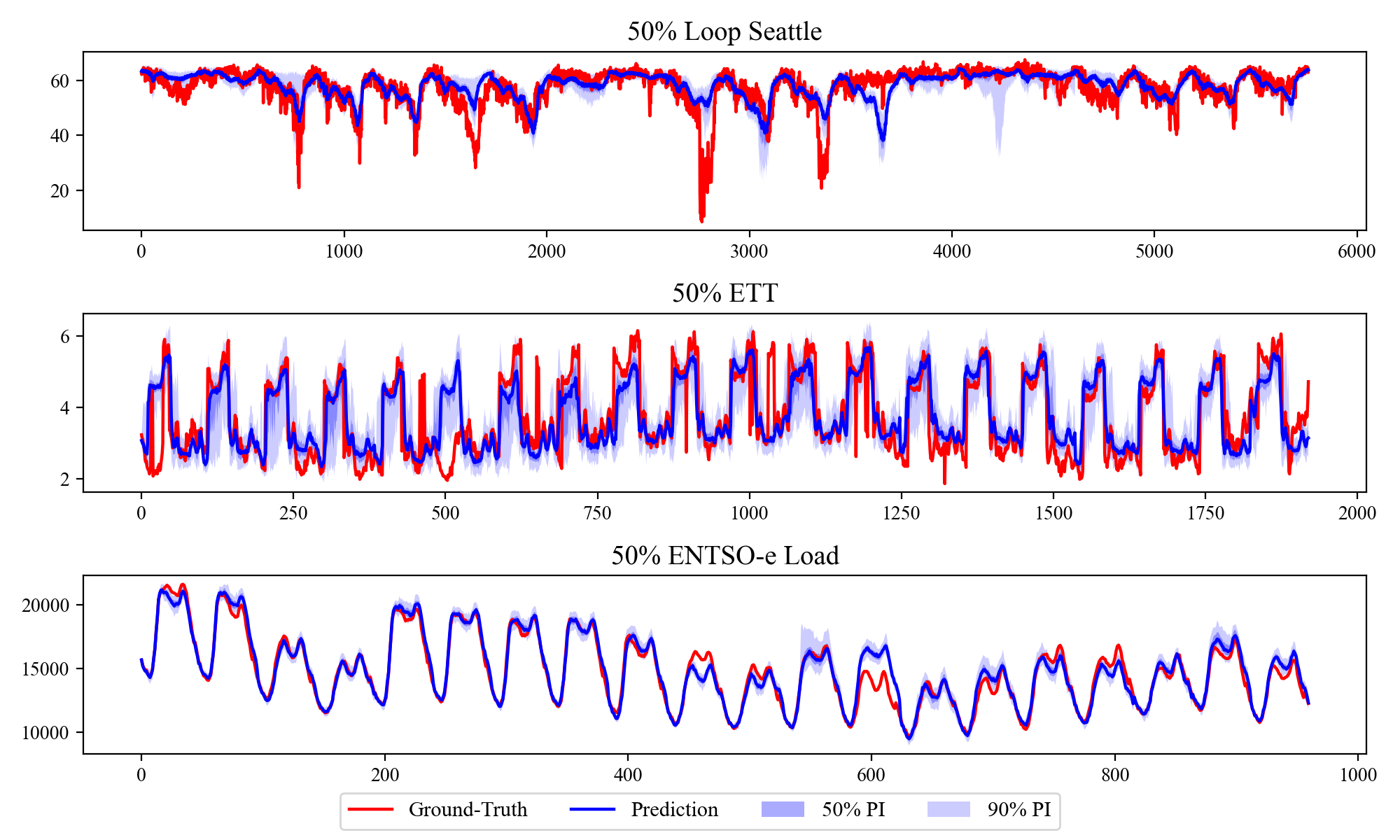}
\vspace{-10pt}
\caption{Visualization of few-shot point forecasts and prediction intervals produced by TimeRFT across 20 testing windows in three datasets.}
\label{fig:pi_showcases}
\end{figure}

\subsection{Parameter Sensitivity Results}
As GRPO-based RFT methods benefit from exploring a larger set of on-policy forecasts, we use the ETT dataset in the 20\% few-shot setting to examine the effect of group size $G$. Figure \ref{fig:group_size_scalability} shows that both TimeGRPO and TimeRFT improve steadily as $G$ increases, with monotonically decreasing MSE and MAE. This suggests that larger groups expose the policy to more diverse on-policy samples outside the training distribution, thereby providing richer exploration signals for learning generalizable temporal patterns. Across all five group sizes, TimeRFT consistently outperforms TimeGRPO, indicating that its gains come not only from larger groups but also from the proposed forecasting-oriented reward design and difficulty-aware data selection.

\begin{table}[ht]
\centering
\caption{Comparison of training overhead on 20\% ETT.}
\vspace{-6pt}
\label{tab:train_time}
\resizebox{0.478\textwidth}{!}{
\begin{tabular}{c|c|ccccc}
\toprule
\multirow{2}{*}{Methods} & \multirow{2}{*}{TimeSFT} & \multicolumn{5}{c}{TimeRFT}           \\ \cmidrule{3-7} 
                         &                          & G=2   & G=4   & G=8   & G=16  & G=24  \\ \midrule
Training Time            & 14min                    & 16min & 18min & 24min & 37min & 50min \\ 
\bottomrule
\end{tabular}
}
\end{table}
\subsection{Training Overhead Analysis}
Table \ref{tab:train_time} reports the training cost of TimeSFT and TimeRFT on the 20\% few-shot ETT setting. TimeRFT introduces additional overhead due to on-policy generation and group-wise optimization, and its cost increases with the group size $G$. Specifically, the training time rises from 16 minutes at $G{=}2$ to 50 minutes at $G{=}24$, compared with 14 minutes for TimeSFT. While this overhead is expected for RL-based exploration, the results show that TimeRFT remains computationally practical at moderate group sizes, especially when balancing training cost and forecasting gains. Improving TimeRFT's training efficiency is an important direction for future work.

\subsection{Data Scalability Analysis}
We evaluate the scalability of the two finetuning paradigms using the univariate Loop Seattle dataset and the multivariate ETT dataset. Figure \ref{fig:data_scalability} shows that prediction errors decrease consistently as more training data becomes available, confirming that TSFM finetuning benefits from additional domain-specific sequences. However, the RFT-based methods are noticeably more sample-efficient: they improve more rapidly in low-data regimes, whereas SFT-based methods adapt more slowly and are more prone to overfitting. Among all methods, TimeRFT consistently achieves the lowest MSE and MAE across data scales and both datasets, and its advantage becomes especially clear when only a small amount of training data is available. This indicates that the proposed forecasting-oriented RFT strategies not only improve forecasting accuracy, but also make TSFM adaptation more scalable under limited supervision. Figure \ref{fig:pi_showcases} further shows the 50\% few-shot forecasts on three datasets, where TimeRFT produces sharpe prediction intervals and faithful future trajectories.

\subsection{Horizon Scaling Results}
\begin{figure}[ht]
\centering
\includegraphics[width=1.0\linewidth]{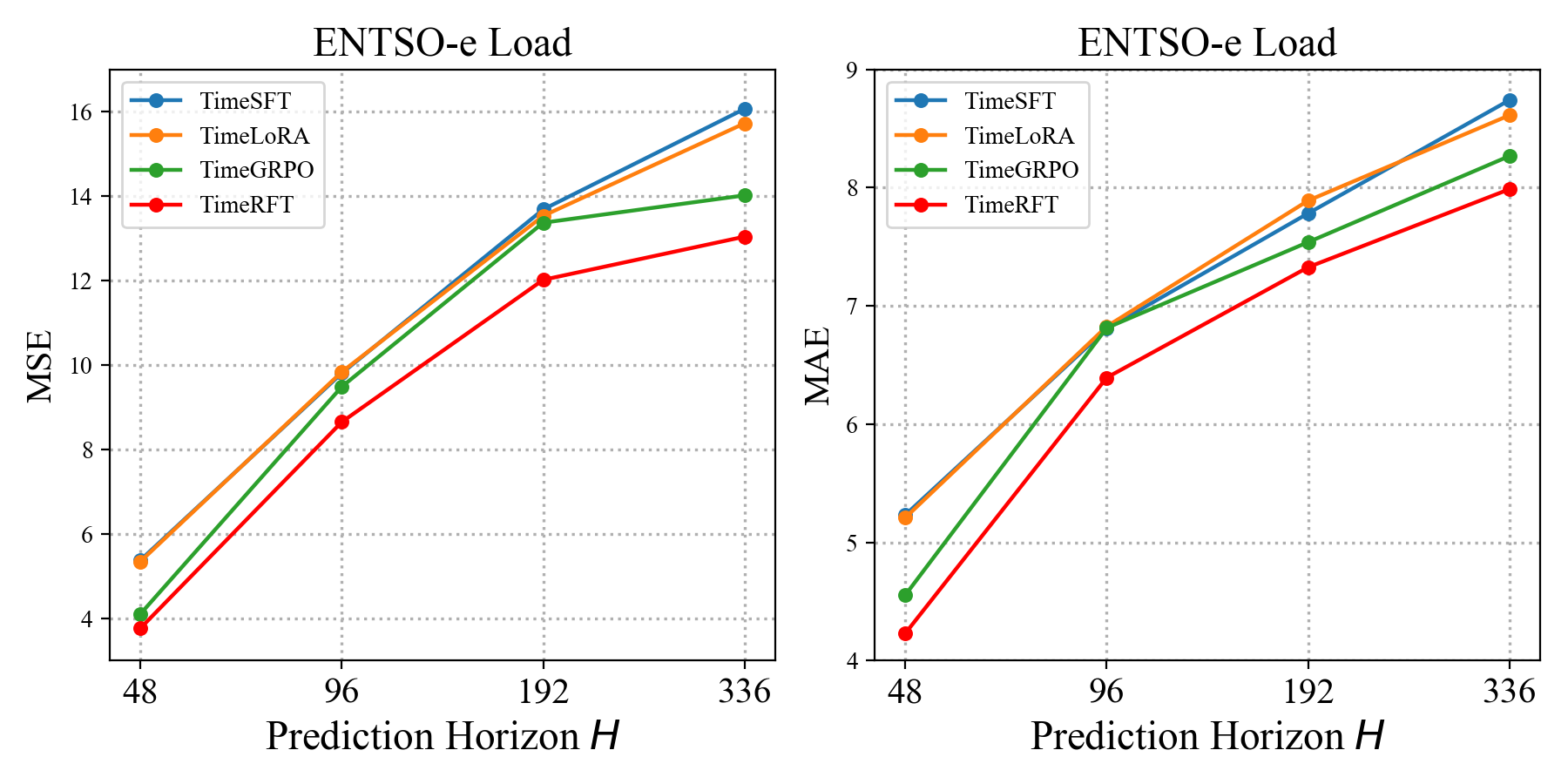}
\vspace{-10pt}
\caption{Comparison on varying prediction horizon $H$.}
\label{fig:horizon_scalability}
\end{figure}

We use the covariate-informed ENTSO-e Load dataset to examine how TimeRFT scales with the prediction horizon, since forecasting from short-term to long-term horizons is a core requirement in power system operation. As shown in Figure \ref{fig:horizon_scalability}, we increase the horizon from $H=48$ to $H=336$ while keeping the lookback length fixed at $L=1008$. As expected, all methods degrade as the horizon grows, reflecting the rising uncertainty and error accumulation in long-range forecasting. Nevertheless, TimeRFT consistently achieves the lowest MSE and MAE across all horizons, suggesting that it learns more transferable temporal dependencies that remain effective for both short-term and long-term prediction.

\end{document}